\documentstyle[preprint,aps,epsfig]{revtex}
\tightenlines
\begin{document}
\draft
\preprint{
\vbox{
\halign{&##\hfil\cr
        & hep-ph/0107309 \cr
        & ANL-HEP-01-057 \cr }}
}
%%%%%%%%%%%%%%%%%%%%%%%%%%%%%%%%%%%%%%%%%%%%%%%%%%%%%%%%%%%%%%%%%%%%%%%%%
%%%%%%%%%%%% Begin Cover Page %%%%%%%%%%%%%%%%%%%%%%%%%%%%%%%%%%%

\title{QCD Factorized Drell-Yan Cross Section \\ 
at Large Transverse Momentum}
\author{Edmond L. Berger$^a$, Jianwei Qiu$^b$ and Xiaofei Zhang$^b$}
\address{
 $^a$Division of High Energy Physics, 
     Argonne National Laboratory \\
     Argonne, Illinois 60439, USA \\
 $^b$Department of Physics and Astronomy,
     Iowa State University \\
     Ames, Iowa 50011, USA}

\date{July 30, 2001}
\maketitle
\begin{abstract}
We derive a new factorization formula in perturbative quantum chromodynamics 
for the Drell-Yan massive lepton-pair cross section as a function of 
the transverse momentum $Q_T$ of the pair.  When $Q_T$ is much larger than 
the pair's invariant mass $Q$, this factorization formula systematically 
resums the logarithmic contributions of the type $\alpha_s^m \ln^m(Q_T^2/Q^2)$ 
to all orders in the strong coupling $\alpha_s$.  When $Q_T\sim Q$, our 
formula yields the same Drell-Yan cross 
section as conventional fixed order QCD perturbation theory.  We show that 
resummation is important when the collision energy $\sqrt{S}$ is large enough 
and $Q_T\gg Q$, and we argue
that perturbative expansions are more stable and reliable in terms of the 
modified factorization formula.     
\end{abstract}
\vspace{0.2in}


%%%%%%%%%%%%%%%%%%%%%%%%%%%%%%%%%%%%%%%%%%%%%%%%%%%%%%%%%%%%%%%%%%%%%%%%%
\section{Introduction}
\label{sec1}

The study of massive lepton-pair production in hadronic collisions (the 
Drell Yan process) has been a valuable pursuit for many years~\cite{DY1}.  
The process is an excellent laboratory for theoretical and experimental 
investigations of 
strong interaction dynamics, and it is a channel for discovery of quarkonium 
states and intermediate vector bosons. In the Drell Yan process, the massive 
lepton-pair is produced via the 
decay of an intermediate virtual photon, $\gamma^*$.  Within the context of 
perturbative quantum chromodynamics (QCD), the Drell-Yan cross section in 
a collision between hadrons $A$ and
$B$, $A(P_A)+B(P_B)\rightarrow \gamma^*(\rightarrow l\bar{l}(Q))+X$,
can be expressed in terms of the cross section for production of an 
unpolarized virtual photon of the same invariant mass~\cite{BGK-DY}, 
\begin{equation}
\frac{d\sigma_{AB\rightarrow \ell^+\ell^-(Q) X}}{dQ^2\,dQ_T^2\,dy}
= \left(\frac{\alpha_{em}}{3\pi Q^2}\right)
  \frac{d\sigma_{AB\rightarrow \gamma^*(Q) X}}{dQ_T^2\,dy}\, .
\label{DY-Vph}
\end{equation}
The variables $Q$, $Q_T$, and $y$ are the invariant mass, transverse momentum, 
and rapidity of the pair.  Symbol $X$ stands for an inclusive sum over final 
states that recoil against the virtual photon. 
An integration has been performed over the angular 
distribution in the lepton-pair rest frame. Because the leptons can be 
detected and measured without restrictions, massive lepton-pair 
production as well as inclusive virtual photon production defined in 
Eq.~(\ref{DY-Vph}) are entirely inclusive.  

Precise knowledge of the gluon parton distribution in nucleons is critical 
for reliable predictions of the signals and backgrounds for many important 
reactions studied at the Fermilab Tevatron and CERN Large Hadron Collider (LHC).  
It was pointed out recently that the transverse momentum distribution of 
massive lepton-pairs produced in hadronic collisions is an advantageous source of 
constraints on the gluon distribution~\cite{BGK-DY}, free from the experimental and 
theoretical complications of photon isolation that beset studies of prompt photon 
production~\cite{BQ-Photon,BGQ-iso}.    Other than the difference between a 
virtual and a real photon, the Drell-Yan process and prompt photon production 
share the same partonic subprocesses.  Similar to prompt photon production, 
the lowest-order virtual photon ``Compton" subprocess: 
$g+q\rightarrow \gamma^*+q$ 
dominates the $Q_T$ distribution when $Q_T > Q/2$, and the next-to-leading 
order contributions preserve the fact that the $Q_T$ distributions are 
dominated by gluon initiated partonic subprocesses~\cite{BGK-DY}. 

If both physically measured quantities $Q$ and $Q_T$ are large, the 
cross section for lepton pairs of invariant mass $Q$ and transverse momentum 
$Q_T$ can be factored systematically in QCD perturbation theory and expressed 
as~\cite{CSS-fac,Brodsky} 
\begin{equation}
\frac{d\sigma_{AB\rightarrow \gamma^*(Q) X}}{dQ_T^2\,dy}
=\sum_{a,b}\int dx_1 \phi_{a/A}(x_1,\mu) 
           \int dx_2 \phi_{b/B}(x_2,\mu)\,
 \frac{d\hat{\sigma}_{ab\rightarrow \gamma^*(Q) X}}{dQ_T^2\,dy}
 (x_1,x_2,Q,Q_T,y;\mu) .  
\label{Vph-fac}
\end{equation}
The sum $\sum_{a,b}$ runs over all parton flavors; $\phi_{a/A}$
and $\phi_{b/B}$ are normal parton distributions; and $\mu$ is 
the renormalization and the factorization scale.  The function 
$d\hat{\sigma}_{ab\rightarrow \gamma^*(Q) X}/dQ_T^2 dy$ in 
Eq.~(\ref{Vph-fac}) represents the short-distance physics of the
collision and is calculable perturbatively in terms of a power
series in $\alpha_s(\mu)$.  The leading order and next-to-leading order 
contributions are available~\cite{BGK-DY,AR-DY}.  The scale $\mu$ is 
of the order of the energy exchange in the reaction, 
$\mu \sim \sqrt{Q^2 + Q_T^2}$.  

There is a phase space penalty associated with the finite mass of 
the virtual photon, and the Drell-Yan factor $\alpha_{em}/(3\pi
Q^2)< 10^{-3}/Q^2$ renders the production rates for massive
lepton pairs small at large values of $Q$ and $Q_T$.  In order to enhance the
Drell-Yan cross section while keeping the dominance of the gluon
initiated subprocesses, it is useful to study lepton pairs with low
invariant mass and relatively large transverse momenta~\cite{BGK-DY}.  With the
large transverse momentum $Q_T$ setting the hard scale of the 
collision, the invariant mass of the virtual photon $Q$ can be small,
as long as the process can be identified experimentally, and the numerical 
value $Q\gg\Lambda_{\rm QCD}$.  For example, the cross section for Drell-Yan
production was measured by the CERN UA1 Collaboration~\cite{UA1-Vph} 
for virtual photon mass $Q\in [2m_\mu, 2.5]$~GeV.  

When $Q_T$ is very different from $Q$ while both are much larger
than $\Lambda_{\rm QCD}$, the calculation of 
massive lepton-pair production becomes a two-scale problem in QCD
perturbation theory.  The corresponding short-distance partonic parts,
calculated in conventional fixed-order QCD perturbation theory,
include potentially large terms proportional to the logarithm of the ratio 
of these two physical scales.  As a result, the higher-order corrections in 
powers of $\alpha_s$ are not necessarily small.  The ratio 
$\sigma^{NLO}/\sigma^{LO}$ 
[$\propto\, \alpha_s\, \times\,$(large logarithms)] can be of 
order 1, and the convergence of the conventional perturbative expansion
in powers of $\alpha_s$ is possibly impaired.  

When $Q_T^2\ll Q^2$, the Drell-Yan (or $W^{\pm}$ and $Z$) 
transverse momentum distributions calculated in fixed-order QCD
perturbation theory are known not to be reliable~\cite{CSS-DYresum}.
After all-orders resummation of the large $\ln(Q^2/Q_T^2)$ terms is 
performed, predictions for the transverse momentum distributions 
become consistent with data for $Q_T^2\ll Q^2$~\cite{RKE-resum,QZ-DYresum}. 
Similarly, when $Q_T^2\gg Q^2$, the region of phase space of interest in 
this paper, the perturbatively calculated short-distance
partonic parts, $d\hat{\sigma}_{ab\rightarrow \gamma^*(Q) X}/dQ_T^2 dy$
in Eq.~(\ref{Vph-fac}), receive one power of the logarithm
$\ln(Q_T^2/Q^2)$ at every order of $\alpha_s$ beyond the leading
order.  At sufficiently large $Q_T$, the coefficients of
the perturbative expansion in $\alpha_s$ will have large logarithmic terms, 
and these high order corrections may not be small.  In order to derive 
reliable QCD predictions, resummation of the logarithmic terms 
$\ln^m(Q_T^2/Q^2)$ must be considered.  

The purpose of this paper is to modify the factorization formula in
Eq.~(\ref{Vph-fac}) so that resummation of the logarithmic contributions 
can be included naturally when $Q_T^2\gg Q^2$.  At the same time, the modified 
factorization formula should remain effectively the same as the conventional 
factorization formula in Eq.~(\ref{Vph-fac}) when $Q_T^2\sim Q^2$.  

In the next section, we review the general structure of the Drell-Yan
cross section, and we identify where the large logarithms arise when
$Q_T^2\gg Q^2\gg \Lambda_{\rm QCD}^2$.  We show that when $Q^2$ is fixed
and $Q^2/Q_T^2\rightarrow 0$, the Drell-Yan cross section behaves 
similarly to the cross section for prompt real photon production~\cite{BGK-DY}.  
The large logarithmic contributions to the Drell-Yan cross
section at high $Q_T$ come from partonic processes that fall into the 
two-stage generic pattern of fragmentation contributions: (1)
short-distance production of a parton of momentum $p_c$, and (2)
fragmentation of this parton into the observed virtual photon.  

In Sec.~\ref{sec3}, we show that the $\ln^m(Q_T^2/Q^2)$ 
logarithmic contributions to the Drell-Yan cross section can be
resummed systematically to all orders in $\alpha_s$.  We demonstrate
that these resummed logarithmic contributions have the same factored
form as those for single hadron production at large transverse momentum,
with the parton-to-hadron fragmentation functions replaced by the
fragmentation functions for a parton to a virtual photon of invariant
mass $Q$.  As for single hadron production, the 
short-distance production of the parton is evaluated at a single hard
scale ($\sim Q_T$), and it is calculable perturbatively in a power series of 
$\alpha_s$.  However, unlike the parton-to-hadron fragmentation
functions, the parton-to-virtual-photon fragmentation functions are
calculable perturbatively if $Q^2\gg\Lambda_{\rm QCD}^2$~\cite{QZ-VPFF}. 

In addition to the logarithmic contributions, the Drell-Yan cross section
includes large {\em non}-logarithmic contributions, in particular, the 
leading order contributions, referred to often as {\em direct} contributions.  
In Sec.~\ref{sec4}, we derive our modified factorization
formula for massive lepton-pair production, Eq.~(\ref{DY-mfac}), in which 
both logarithmic and non-logarithmic contributions are fully included.  
This modified factorization formula agrees with the conventional fixed-order 
QCD expression when $Q_T^2\sim Q^2$ (or when the logarithmic
contributions are less important).  We show that at
the next-to-leading order, the difference between the modified
factorization formula and the conventional factorization formula is
completely determined by QCD evolution of the virtual photon fragmentation 
functions.  Our modified factorization formalism reorganizes the
{\it single} perturbative expansion of conventional QCD factorization into 
{\it two} perturbative expansions plus the perturbatively calculated
parton-to-virtual photon fragmentation functions.  The main advantage of
this reorganization is that the new perturbative expansions are 
evaluated at a single hard scale and are free of large logarithmic terms 
for $Q_T \ge Q$.

In Sec.~\ref{sec5}, we present our predictions for the cross sections 
for massive lepton-pair production at energies of interest for experiments 
at the Fermilab Tevatron, Brookhaven's RHIC, and the CERN Large Hadron 
Collider.  We include both leading order and next-to-leading order direct  
short-distance contributions and the resummed logarithmic contributions. The 
resummed large logarithmic contributions change the shape of the predicted 
$Q_T$-spectrum of the Drell-Yan cross section but, 
at the order in perturbation theory at which we work, they  
have only a modest effect on the normalization.  
We confirm that after the large logarithmic terms are resummed to all 
orders in $\alpha_s$ the Drell-Yan cross section at large $Q_T$ remains an 
excellent source of contraints on the gluon parton density.  Our conclusions are 
summarized in Sec.~\ref{sec6}.

%%%%%%%%%%%%%%%%%%%%%%%%%%%%%%%%%%%%%%%%%%%%%%%%%%%%%%%%%%%%%%%%%%%%%%%%%
\section{Massive Lepton Pair Production at Fixed Order} 
\label{sec2}

In hadronic collisions, massive lepton pair production
proceeds through partonic hard-scattering processes involving
initial-state quarks and gluons.  If the lepton pair's invariant mass
$Q$ and its transverse momentum $Q_T$ are both much larger than 
$\Lambda_{\rm QCD}$, the partonic hard-scattering at a distance scale
between $O(1/Q)$ and $O(1/Q_T)$ can be systematically factored from
the physics at the scale of hadron wave functions, $O(1/\Lambda_{\rm
QCD})$. In this situation, the cross section can be
expressed in the factored form of Eq.~(\ref{Vph-fac}).  
Corrections to the expression in Eq.~(\ref{Vph-fac}) are
suppressed by 
powers of $\Lambda_{\rm QCD}^2/Q^2$ or $\Lambda_{\rm QCD}^2/Q_T^2$.  
The predictive power of 
Eq.~(\ref{Vph-fac}) relies on the {\it universality} of the parton
distributions and the {\it reliability} of the partonic cross
sections.  

The short-distance partonic cross sections,
$d\hat{\sigma}_{ab\rightarrow \gamma^*(Q) X}/dQ_T^2 dy$ in
Eq.~(\ref{Vph-fac}), can be calculated in principle order by order in
QCD perturbation theory in a power series of the strong coupling
$\alpha_s$,    
\begin{equation}
\frac{d\hat{\sigma}_{ab\rightarrow \gamma^*(Q) X}}{dQ_T^2\,dy}
=\sum_{n=0} H^{(n)}_{ab\rightarrow \gamma^* X}(x_1,x_2,Q,Q_T,y;\mu)
\left(\frac{\alpha_s(\mu)}{2\pi}\right)^n\, .
\label{Vph-Hab}
\end{equation}
The reliability of QCD perturbative calculations depends on
the behavior of the coefficient functions 
$H^{(n)}_{ab\rightarrow \gamma^* X}$ in Eq.~(\ref{Vph-Hab}).    

At lowest-order, $O(\alpha_s^0)$, the only partonic subprocess for
virtual photon production is $q+\bar{q}\rightarrow \gamma^*$.  The 
incoming
partons are assumed to be collinear to their respective incoming
hadrons if power suppressed corrections are neglected.  Therefore,  
the lowest order coefficient function, 
$H^{(0)}_{q\bar{q}\rightarrow \gamma^* X} \propto \delta(Q_T)$, 
vanishes if $Q_T\neq 0$.  

\begin{figure}
\begin{center}
\epsfig{figure=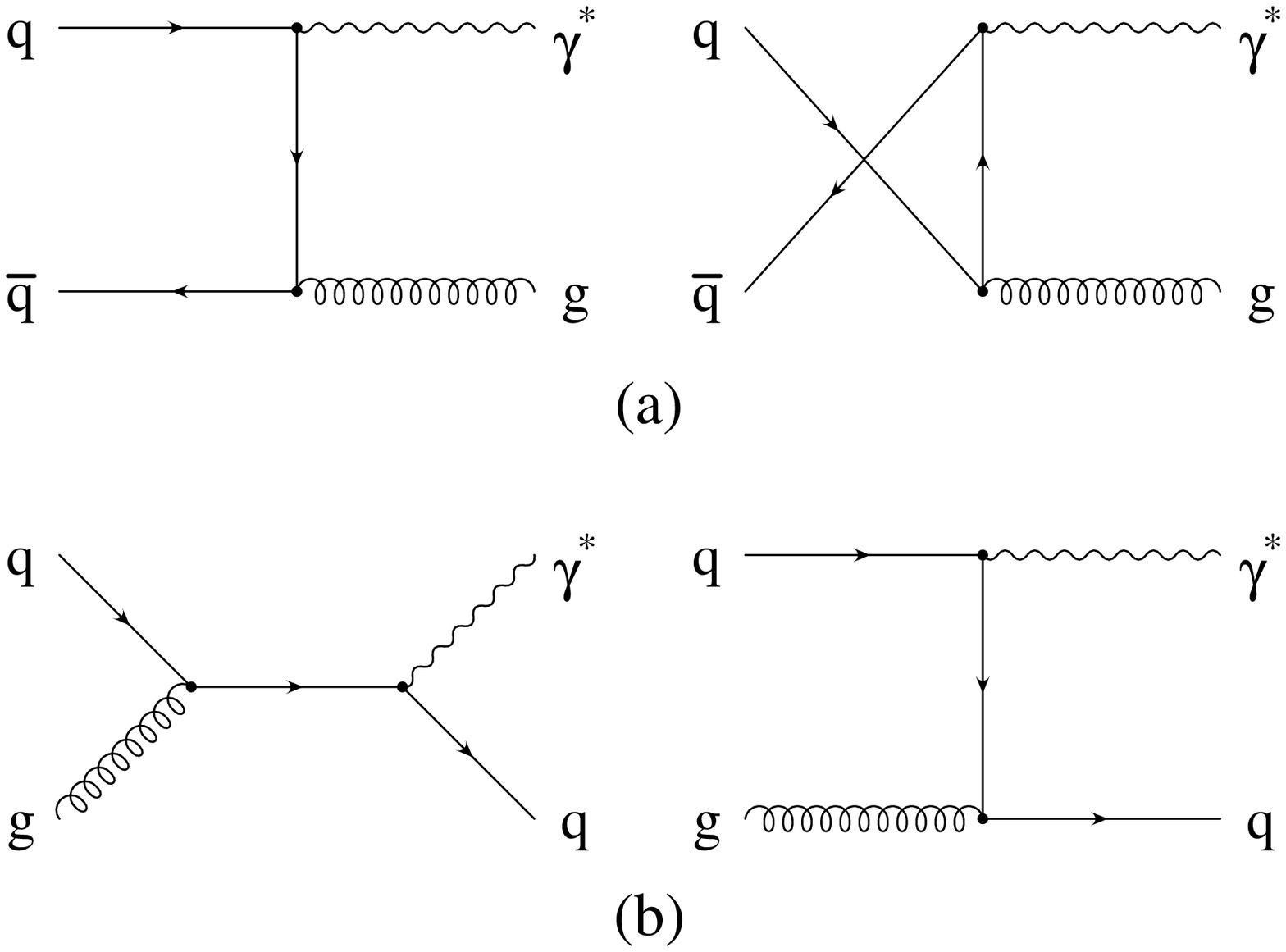,width=4.5in}
\end{center}
\caption{Feynman diagrams for the LO contribution to the Drell-Yan
cross section: (a) quark-antiquark annihilation
$q+\bar{q}\rightarrow\gamma^*+g$, and (b) Compton 
$g+q\rightarrow\gamma^*+q$ subprocesses.}
\label{fig1}
\end{figure}

At order $O(\alpha_s)$, both quark-antiquark annihilation,
$q+\bar{q}\rightarrow \gamma^* + g$, and ``Compton'',
$g+q\rightarrow\gamma^*+q$, subprocesses contribute to 
the Drell-Yan cross section, with the recoil of the final-state parton
balancing the transverse momentum of the lepton pair.  These partonic
subprocesses, known as the $2\rightarrow 2$ subprocesses, are shown
in Figs.~\ref{fig1}(a) and (b).  They are often referred as the
leading order (LO) contributions to the Drell-Yan cross section with
finite transverse momentum.  The corresponding leading order coefficient 
functions are 
\begin{equation}
H^{(1)}_{ab\rightarrow \gamma^* X}
= e_q^2\, \frac{\pi}{2 x_1 x_2 S}
\left|\frac{1}{g_s} \overline{M}_{ab\rightarrow\gamma^* X} \right|^2\,
\left(8\pi^2\right)\,
\delta\left((x_1P_A+x_2P_B-Q)^2\right) 
\left(\frac{\alpha_{em}}{2\pi}\right)\, .
\label{H-LO}
\end{equation}
The incoming parton flavors ``$ab$'' can be either $q\bar{q}$
for the quark-antiquark annihilation or $gq$ for the Compton
subprocess, and $e_q$ is quark's fractional charge.  
In Eq.~(\ref{H-LO}), $g_s$ and $\alpha_{em}$ are
the strong coupling constant and the fine structure constant of QED,
respectively; $P_A$ and $P_B$ are the momenta of the colliding hadrons, 
and $S=(P_A+P_B)^2$ is the square of the total collision energy. The 
expressions   
$|\overline{M}_{ab\rightarrow\gamma^* X}|^2$ in Eq.~(\ref{H-LO}) are
the squares of the matrix elements of the partonic subprocesses shown in
Fig.~\ref{fig1}, summed (averaged) over the colors and spins of the 
final-state (initial-state) partons. They are available in 
Ref.~\cite{Owens-RMP}.
As long as $Q_T$ is large, the LO coefficient functions in
Eq.~(\ref{H-LO}) are well-behaved, even when $Q^2\rightarrow 0$.

The calculation of the perturbative coefficient functions at order
$O(\alpha_s^2)$, known as the next-to-leading order (NLO)
contribution, involve all $2\rightarrow 3$ partonic subprocesses with
the virtual photon in the 
final-state as well as the $2\rightarrow 2$ diagrams in
Fig.~\ref{fig1} with one-loop corrections.
After renormalization, the loop momentum integrations for the
$2\rightarrow 2$ diagrams at order $O(\alpha_s^2)$ yield 
renormalization scale ($\mu$) dependence and logarithmic terms in the
coefficient functions $H^{(2)}_{ab\rightarrow\gamma^* X}$. 
Integration over the phase space of the     
extra parton in the final-state of the $2\rightarrow 3$
subprocesses leads to a collinear
divergence when this parton is collinear to either incoming parton.
QCD factorization and subtraction of the collinear divergence results 
in factorization scale ($\mu_f$) dependence and logarithmic terms 
in the coefficient functions at this order.  Consequently, the
coefficient functions $H^{(2)}_{ab\rightarrow\gamma^* X}$ display 
logarithmic dependence on the ratios of the following momentum scales:
$\mu$, $\mu_f$, $Q_T$, and $Q$ \cite{BGK-DY,AR-DY}.

Since we are interested in identifying the logarithms of the ratio
$Q_T^2/Q^2$, with $Q_T^2\gg Q^2$, we concentrate on the part of the
Drell-Yan cross section that diverges when $\ln(Q_T^2/Q^2)\rightarrow
\infty$ with $Q^2\gg\Lambda_{\rm QCD}^2$.  According to the QCD
factorization theorem, the perturbatively calculated partonic cross
sections $d\hat{\sigma}_{ab\rightarrow\gamma^*(Q) X}/dQ_T^2 dy$ in
Eq.~(\ref{Vph-fac}) should be analytic functions of $Q_T^2$ and $Q^2$.
Therefore, we expect the logarithmic behavior of the 
Drell-Yan cross section 
as $\ln(Q_T^2/Q^2)\rightarrow\infty$ with $Q^2\gg\Lambda_{\rm QCD}^2$
fixed to be connected closely to logarithmic divergences associated with 
the massless photon ($Q^2=0$) in the case of prompt real photon production.  

Other than the non-vanishing invariant mass, production of a virtual
photon and a real photon share the same parton-level Feynman diagrams.
However, the QCD factorization formula for production of a real photon
($Q^2=0$) is different from that in Eq.~(\ref{Vph-fac}),
\begin{equation}
\frac{d\sigma_{AB\rightarrow \gamma X}}{dQ_T^2\,dy}
=\sum_{a,b}\int dx_1 \phi_{a/A}(x_1,\mu) 
           \int dx_2 \phi_{b/B}(x_2,\mu)
\left[ 
\frac{d\hat{\sigma}_{ab\rightarrow \gamma X}^{(Dir)}}{dQ_T^2\,dy}
+
\frac{d\hat{\sigma}_{ab\rightarrow \gamma X}^{(F)}}{dQ_T^2\,dy}
\right] , 
\label{Ph-fac}
\end{equation}
where $d\hat{\sigma}_{ab\rightarrow \gamma X}^{(Dir)}/dQ_T^2 dy$
represents the direct production of the real photon at a
short-distance scale of $O(1/Q_T)$; and 
\begin{equation}
\frac{d\hat{\sigma}_{ab\rightarrow \gamma X}^{(F)}}{dQ_T^2\,dy}
= \sum_c \int \frac{dz}{z^2}\,
\frac{d\hat{\sigma}_{ab\rightarrow c X}}{dp_{c_T}^2\,dy}
\left(p_c=\frac{Q}{z},\mu_F^2\right)
D_{c\rightarrow\gamma X}(z,\mu_F^2)
\label{Ph-fac-F}
\end{equation}
is the fragmentation contribution to the prompt photon cross
section.  The function $d\hat{\sigma}_{ab\rightarrow c X}/dp_{c_T}^2 dy$ in
Eq.~(\ref{Ph-fac-F}) represents short-distance production of a parton
of flavor $c$, and $D_{c\rightarrow\gamma X}(z,\mu_F^2)$ is a
fragmentation function for parton $c$ to fragment into a real
photon with photon momentum $Q=z\,p_c$; $\mu_F$ is the fragmentation scale. 

The fragmentation contribution arises because there are collinear 
singularities associated with the region of phase space in which 
the real photon is parallel to one or more of the final-state 
partons~\cite{Owens-RMP,Aurenche-photon}.  Because of the real photon 
is massless, the 
parent parton, which fragments into the real
photon and other collinear partons, can propagate for a long time.
Consequently, quantum interference between the production of the
parent parton and the physics associated with the fragmentation (or
decay) of the parton is suppressed.  Therefore, the
fragmentation contribution to prompt photon production can be
further factored as in Eq.~(\ref{Ph-fac-F}).  Because the 
transverse momentum $Q_T$ is large, all logarithmic collinear divergences
associated with the massless photon arise from final-state partons that 
are parallel to the observed real photon.  Such logarithmic
divergences are all absorbed into the fragmentation functions.  These
functions are nonperturbative in nature.   

Unlike prompt photon production, it is not necessary to introduce 
fragmentation functions to absorb final-state collinear
singularities for the Drell-Yan cross section.  Because the 
photon is off-shell, its large invariant mass $Q$ regulates the
singularity.  This finite mass regularization leads to a logarithmic
dependence of the Drell-Yan cross section on the invariant mass of the
virtual photon $Q$.  If $Q$ is large enough and $Q_T$, the only 
other physically observed momentum scale, is not too large, the
logarithmic terms $\ln^m(Q_T^2/Q^2)$ are small, and no resummation of the
logarithms is necessary for a reliable prediction of the cross section.

When $Q^2$ is chosen to be small, so as to enhance the Drell-Yan cross
section, and when the collision energy $\sqrt{S}$ and $Q_T$ become 
large, it is necessary to examine the size of the final-state logarithmic 
contributions and ascertain whether resummation of these logarithmic 
terms is warranted.

\begin{figure}
\begin{center}
\epsfig{figure=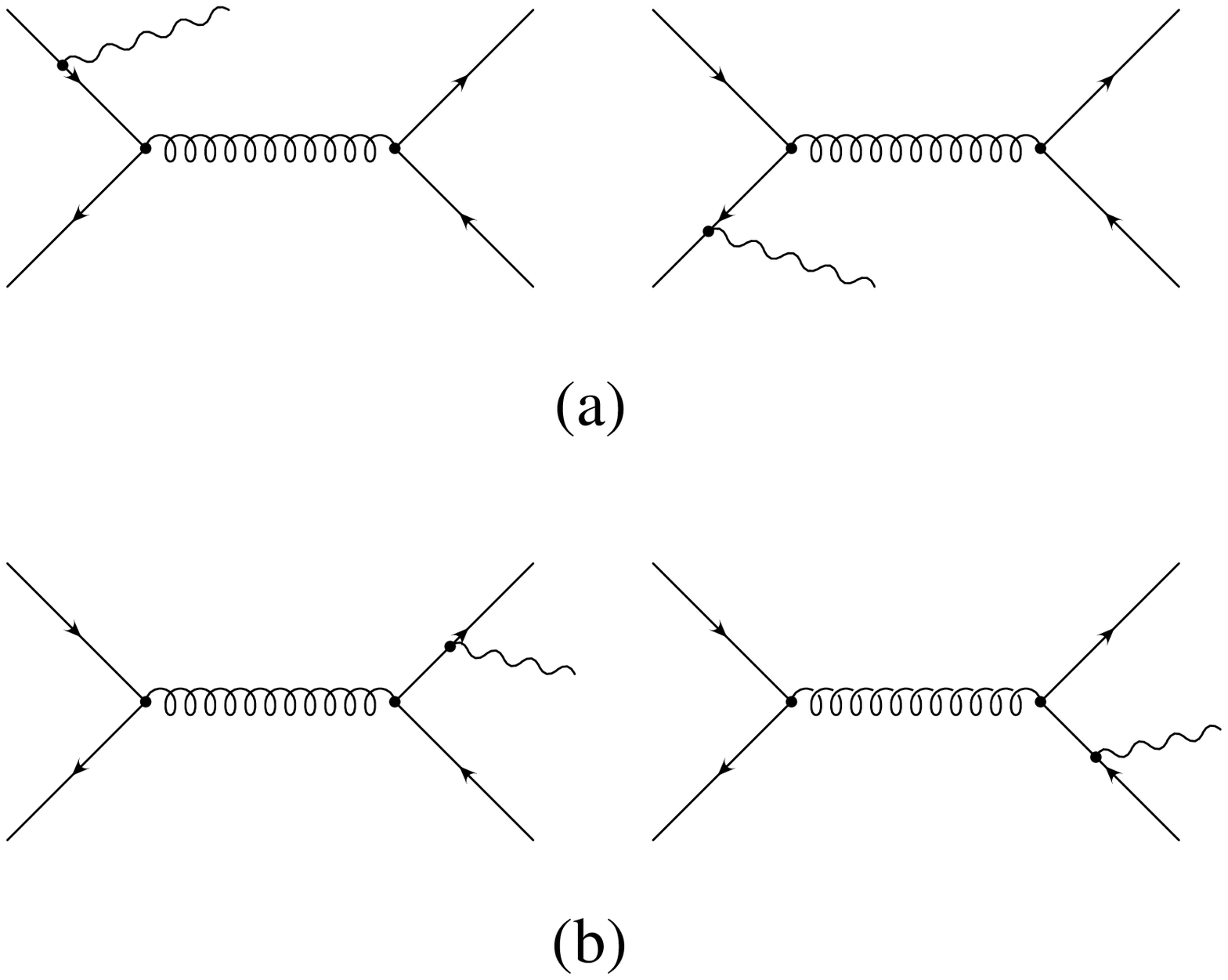,width=4.5in}
\end{center}
\caption{Feynman diagrams that illustrate situations in which the
photon can become collinear to a quark in (a) the initial-state and (b) 
the final-state.} 
\label{fig2}
\end{figure}

The explicit form of NLO contributions to the
Drell-Yan cross section from a $2\rightarrow 3$
partonic subprocess $q+\bar{q}\rightarrow\gamma^*(Q)+q'+\bar{q}'$
\cite{BGK-DY} provides an example of the logarithmic terms under discussion. 
The relevant Feynman diagrams are shown in Fig.~\ref{fig2}(a) and
\ref{fig2}(b) for initial- and final-state photon radiation.  Since we
are interested mainly in the structure of the final state in the
region where a photon becomes collinear to a quark, this subprocess is
typical of the generic $2\rightarrow 3$ subprocess in
Fig.~\ref{fig3}, and conclusions drawn from it can applied to 
other subprocesses such as $qg$ and $qq'$ scattering.  For the phase
space integrals of the diagrams in Fig.~\ref{fig2}, we use the exact
forms in Ref.~\cite{EMP-DY}. The logarithmic contributions have 
the form 
$$
\ln\left[\frac{s+Q^2-s_2+\lambda}{s+Q^2-s_2-\lambda}\right]\,
g(s,t,u,Q^2) .
$$
The function $g(s,t,u,Q^2)$ is given in Ref.~\cite{BGK-DY}.  
It is well-behaved as $Q^2\rightarrow 0$.  The parton-level Mandelstam 
variables $s,t,u$ are defined as 
$s=(p_1+p_2)^2$, $t=(p_1-Q)^2$, and $u=(p_2-Q)^2$.  The function
$\lambda=\sqrt{(t+u)^2-4Q^2 s_2}$, with    
$s_2\equiv (p_1+p_2-Q)^2 = s + t + u - Q^2$; $s_2$ is the square of the 
invariant mass of the two final-state partons that recoil against $Q$.

In the limit $Q^2\ll |t+u|$, the generic logarithm from the
splitting of a quark into a photon can be approximated as
\begin{eqnarray}
\ln\left[\frac{s+Q^2-s_2+\lambda}{s+Q^2-s_2-\lambda}\right]
&\rightarrow & 
\ln\left[\frac{2Q^2(s-Q^2)/(-(t+u))}{2(-(t+u))}\right]
\nonumber \\
&\rightarrow & 
- \ln\left[\frac{-(t+u)}{Q^2}\right]\, .
\label{log-M}
\end{eqnarray}
To derive the second line we use $Q^2\ll |t+u|$ and $s\sim
|t+u|$.  Since $|t+u| \sim O(Q_T^2+Q^2)$, the logarithm takes the form
$\ln[(Q_T^2+Q^2)/Q^2]\sim \ln[Q_T^2/Q^2]$, and the limit of $Q^2\ll
|t+u|$ is effectively the same as $Q^2\ll Q_T^2$.  The logarithmic
contributions to the Drell-Yan cross section become more important
when $Q^2$ is chosen to be small.  

In principle, we can resum these large but finite logarithms into 
parton-to-virtual-photon fragmentation functions, just as 
the logarithmic divergences in real photon production are resummed 
into real-photon fragmentation functions.  Because the logarithms in 
the Drell-Yan case are finite, and $Q^2$ is much larger than $\Lambda_{\rm
QCD}^2$, the parton-to-virtual-photon fragmentation functions should
be calculable perturbatively~\cite{QZ-VPFF}.

%%%%%%%%%%%%%%%%%%%%%%%%%%%%%%%%%%%%%%%%%%%%%%%%%%%%%%%%%%%%%%%%%%%%%%%%%
\section{Fragmentation Contributions to the Drell-Yan Cross Section}
\label{sec3}

In this section, we derive the resummed logarithmic contributions to 
the Drell-Yan cross section when $Q_T^2\gg Q^2$.  Since we are
interested mainly in the cross section at 
low $Q^2$, we ignore contributions from the intermediate vector boson 
$Z$.

%========================================================================
\subsection{Logarithmic Contributions to the Drell-Yan Cross Section at
Large $Q_T$}
\label{sec3a}

As demonstrated in the last section, the Drell-Yan cross section
at large transverse momentum $Q_T$ receives potentially large logarithmic 
terms $\ln^m(Q_T^2/Q^2)$ from the part of phase space in which 
the virtual photon is
almost collinear to one or more final-state partons.  Since at
least one final-state parton is needed to balance the virtual photon's
transverse momentum, the logarithmic contributions can arise only
at NLO and beyond.  

Because of the logarithms, the coefficient functions 
in Eq.~(\ref{Vph-Hab}) might be large, and
resummation of the logarithmic contributions might be needed.
As long as $Q^2$ is much larger than $\Lambda_{\rm QCD}^2$, all
coefficient functions in Eq.~(\ref{Vph-Hab}) are calculable in principle
order by order in QCD perturbation theory.  Therefore,
resummation of the logarithms $\ln^m(Q_T^2/Q^2)$ is actually a
{\it reorganization} of the perturbative expansion in
Eq.~(\ref{Vph-Hab}), such that all coefficient functions in the
reorganized perturbative expansions are evaluated at a single hard
scale and free of any large logarithms.    

The energy exchange in the hard
collision is of the order of $\sqrt{Q_T^2+Q^2}\approx Q_T
+O(Q^2/Q_T^2)$.  When $Q_T^2\gg Q^2$, the partonic hard collision 
should not be sensitive to the scale $Q^2$ at which the virtual photon
is produced.  If we neglect power corrections of the order
$Q^2/Q_T^2$, the partonic hard collisions are effectively independent
of $Q^2$, except for the logarithmic dependence $\ln^m(Q_T^2/Q^2)$
from the final-state bremsstrahlung production of the virtual photon. 
Therefore, other than the appearance of the virtual photon's mass $Q$ 
to regulate the
final-state collinear divergences, the potentially large logarithmic
contributions at large $Q_T$ have the
same structure as the fragmentation contributions to prompt real 
photon production.  They can be separated into two stages: 
(1) production of a parton of momentum $p_c$ at a very short-distance
($\sim 1/p_{c_T}\sim 1/Q_T$), and (2) production of the lepton-pair
via a virtual photon of invariant mass $Q$ through bremsstrahlung
(or fragmentation) from the parton produced at the first
stage.  This two-stage production, shown in Fig.~\ref{fig3} or in
general in Fig.~\ref{fig4}, shares the same generic pattern of the  
fragmentation production of a single particle (e.g., a hadron of mass
$M_h$ or a real photon) at large transverse momentum $Q_T$.
If we neglect the power suppressed quantum interference between these
two stages, the fragmentation (or bremsstrahlung) contributions 
should have the same general factored form that is present 
in single hadron or prompt photon production 
\cite{QZ-VPFF},  
\begin{equation}
\frac{d\hat{\sigma}_{ab\rightarrow\gamma^*(Q) X}^{(F)}}{dQ_T^2 dy}
= \sum_{c} \int \frac{dz}{z^2}\, \left[
  \frac{d\hat{\sigma}_{ab\rightarrow c X}^{(F)}}{dp_{c_T}^2\,dy}
  \left(x_1,x_2,p_c=\hat{Q}/z;\mu_F\right)  \right]
  D_{c\rightarrow \gamma^* X}(z,\mu_F^2;Q^2) .
\label{DY-F}
\end{equation}
Superscript $(F)$ indicates the fragmentation contribution; the sum 
$\sum_c$ runs over all parton flavors; and $\mu_F$ is the
fragmentation scale defined below.  The four vector $\hat{Q}^\mu$ in
Eq.~(\ref{DY-F}) is defined to be $Q^\mu$ but with $Q^2$ set to be zero. 

The partonic cross sections, 
$d\hat{\sigma}_{ab\rightarrow c X}/dp_{c_T}^2\,dy$ in Eq.~(\ref{DY-F}),
represent the inclusive production of a parton of flavor $c$.  These partonic
cross sections are evaluated at a single hard scale $p_{c_T} \sim
Q_T$.  Exact perturbative expressions for the partonic cross
sections depend on how the fragmentation functions
$D_{c\rightarrow\gamma^* X}(z,\mu_F^2;Q^2)$ are defined (or depend on
the choices of the scheme) \cite{QZ-VPFF}.  

As for single hadron (or prompt photon) production, all large
logarithms from the integration over the distance scale from 
$O(1/p_{c_T})$ (or $O(1/Q_T)$) to $O(1/Q)$, where the virtual photon
is produced, are resummed into the fragmentation functions.  The main
difference for production of a virtual photon, a hadron, or a real
photon is the difference in the fragmentation functions.   

%========================================================================
\subsection{Parton-to-Virtual-Photon Fragmentation Functions}
\label{sec3b}

The fragmentation function for a parton of flavor $c$
into a virtual photon of invariant mass Q is defined as a probability
density for finding the virtual photon with fraction $z$ of the parent 
parton's momentum.
Because of the universality of fragmentation functions,
virtual photon fragmentation functions can be derived in a
process independent way \cite{QZ-VPFF} or extracted from a specific
physical process.  In this subsection, we derive the virtual photon
fragmentation functions by resumming the leading logarithmic
contributions to the Drell-Yan cross section at large transverse
momentum.   

In order to resum the large logarithmic contributions to the Drell-Yan
cross section to all orders and to extract the virtual photon
fragmentation functions, we must identify all sources and the
pattern of these logarithmic contributions. 
Consider a non-singlet fragmentation of a quark into a
virtual photon, shown in
Fig.~\ref{fig5}.  The virtual photon is produced from 
bremsstrahlung of a quark (or an antiquark).  The quark (or antiquark)
itself is produced either in the hard collision, as shown in
Fig.~\ref{fig5}(a), or from the fragmentation of another quark, as
shown in Fig.~\ref{fig5}(b) or Fig.~\ref{fig5}(c).  In a physical
gauge (such as the light-cone gauge), the leading large logarithmic
contributions of the subprocesses in Fig.~\ref{fig5} come from the part 
of phase space in which the daughter quark's invariant mass $k_i^2$ is 
much smaller than that of the parent quark $k_{i+1}^2$.  In this 
situation, we can neglect all contributions suppressed by
powers of $k_i^2/k_{i+1}^2$. The ``decay'' rate (or the fragmentation
function) of the parent quark into a daughter quark plus massless
partons is dominated by logarithmic contributions that are
proportional to $\ln(k_{i+1}^2/k_i^2)$.  The leading logarithmic 
contributions from the general fragmentation diagram in Fig.~\ref{fig4} 
arise from the region of 
strong ordering in the invariant masses of the fragmentation
partons in Fig.~\ref{fig5}: $p_c^2\gg ... k_{i+1}^2\gg k_i^2\gg
... \gg k_0^2$ \cite{QZ-VPFF}.    

With the strong ordering approximation, the leading logarithmic
contributions from the non-singlet fragmentation in Fig.~\ref{fig5} can be
factored into the hard production of a quark of momentum
$p_c$ convoluted with a sum of all-orders 
fragmentation ladder diagrams, as shown in Fig.~\ref{fig6}.  By
comparing the factored cross section in Eq.~(\ref{DY-F}) with the
factored expression in Fig.~\ref{fig6}, and summing the ladder
diagrams to all orders, one obtains the leading contributions to the
non-singlet quark-to-virtual-photon fragmentation functions
\cite{QZ-VPFF},  
\begin{eqnarray}
D_{q\rightarrow \gamma^* X}^{NS}(z,p_c^2;Q^2) 
&=& \int_{Q^2/z}^{p_c^2}\,
    \frac{dk_0^2}{k_0^2} \left( 
    \frac{\alpha_{em}}{2\pi}
    \gamma_{q\rightarrow \gamma^*}^{(0)}(z,k_0^2;Q^2)\right)  
\nonumber \\
&+& \sum_{n=1}^{\infty}
    \left[ \prod_{i=1}^{n}
     \int_{Q^2/z}^{k_{i+1}^2}\,
     \frac{dk_i^2}{k_i^2} \left(
       \frac{\alpha_s}{2\pi}
       \int_{z_{i+1}}^1 \frac{dz_i}{z_i}
        P_{q\rightarrow q}^{(0)}(\frac{z_{i+1}}{z_i})\right)
    \right] 
\nonumber \\
&\ & {\hskip 0.2in} \times
    \int_{Q^2/z}^{k_1^2}\,
    \frac{dk_0^2}{k_0^2} \left( 
    \frac{\alpha_{em}}{2\pi}
    \gamma_{q\rightarrow \gamma^*}^{(0)}(z_1,k_0^2;Q^2)\right) .
\label{Dq-NS-sum}
\end{eqnarray}
The superscript ``$NS$'' represents the non-singlet
contribution; the upper limit of integration is $k_{n+1}^2 = p_c^2$; and  
the lower limit of integration, $Q^2/z$, is the mass threshold (or
minimum invariant mass) for the quark to produce the virtual photon of
invariant mass $Q$ \cite{QZ-VPFF}.  The
leading-order quark-to-quark splitting function $P_{q\rightarrow
q}^{(0)}(z)$ is the same as the leading-order quark-to-quark splitting
function of the DGLAP evolution equations \cite{CQ-evo}.  In deriving
Eq.~(\ref{Dq-NS-sum}), we include the diagrams with quark wave
function renormalization in addition to the ladder diagrams shown in
Fig.~\ref{fig6}.  The function $\gamma_{q\rightarrow
\gamma^*}^{(0)}(z,k_0^2;Q^2)$ in Eq.~(\ref{Dq-NS-sum}) is the lowest 
order QED splitting function for a quark to fragment into a virtual
photon \cite{QZ-VPFF,BL-frag}, defined below. 

Since the invariant mass of the parent quark, $p_c^2$, can be very
large in high energy collisions, the resummed logarithmic
contributions, given by the second term on the
right-hand-side of Eq.~(\ref{Dq-NS-sum}), can be important for the
Drell-Yan cross section at low $Q^2$.  From Eq.~(\ref{DY-F}) and the
fact that the partonic hard parts are evaluated at a single hard scale
$\sim p_{c_T}\sim Q_T$, we conclude that all leading large logarithmic 
contributions at high $Q_T$ are included in the
virtual photon fragmentation functions.  Resummation of the large
logarithmic contributions is equivalent to the derivation of
the virtual photon 
fragmentation functions in Eq.~(\ref{Dq-NS-sum}).  Unlike the
parton-to-hadron fragmentation functions, the virtual photon
fragmentation functions in Eq.~(\ref{Dq-NS-sum}) have no dependence on
any non-perturbative momentum scale. The virtual
photon's non-vanishing invariant mass removes the final-state
collinear singularities that appear in the parton-to-hadron (or
real photon) fragmentation functions.  Therefore, all
parton-to-virtual-photon fragmentation functions should be free of 
collinear singularities, and they are calculable in principle perturbatively
to all orders in $\alpha_s$ \cite{QZ-VPFF}.  

By re-organizing the second term on the right-hand-side of
Eq.~(\ref{Dq-NS-sum}), one may derive an integral equation for the
non-singlet quark-to-virtual-photon fragmentation function,
\begin{eqnarray}
D_{q\rightarrow \gamma^* X}^{NS}(z,p_c^2;Q^2) 
&=& \int_{Q^2/z}^{p_c^2}\,
    \frac{dk_0^2}{k_0^2} \left( 
      \frac{\alpha_{em}}{2\pi}
      \gamma_{q\rightarrow \gamma^*}^{(0)}(z,k_0^2;Q^2)\right)  
\nonumber \\
&+& \int_{Q^2/z}^{p_c^2}\,
    \frac{dk^2}{k^2} \left(
      \frac{\alpha_s}{2\pi}
      \int_{z}^1 \frac{dz'}{z'}
        P_{q\rightarrow q}^{(0)}(\frac{z}{z'})\right)
    D_{q\rightarrow \gamma^* X}^{NS}(z',k^2;Q)\, .
\label{Dq-NS-intg}
\end{eqnarray}
Carrying out the sum over all-orders logarithmic contributions
in Eq.~(\ref{Dq-NS-sum}) is equivalent to solving the integral
equation in Eq.~(\ref{Dq-NS-intg}).  

We introduce fragmentation scale $\mu_F$ and let $\mu_F^2\equiv p_c^2$,
the square of the invariant mass of the parent quark in Eq.~(\ref{Dq-NS-intg}).
One can derive an evolution equation by applying
$\mu_F^2 d/d\mu_F^2$ to both sides of Eq.~(\ref{Dq-NS-intg}),  
\begin{eqnarray}
\mu_F^2\frac{d}{d\mu_F^2}\,
D_{q\rightarrow \gamma^* X}^{NS}(z,\mu_F^2;Q^2) 
&=& \frac{\alpha_{em}}{2\pi}
    \gamma_{q\rightarrow \gamma^*}^{(0)}(z,\mu_F^2;Q^2)
\nonumber \\
&+& \frac{\alpha_s}{2\pi}
    \int_{z}^1 \frac{dz'}{z'}
      P_{q\rightarrow q}^{(0)}(\frac{z}{z'})
      D_{q\rightarrow \gamma^* X}^{NS}(z',\mu_F^2;Q^2)\, .
\label{Dq-NS-evo}
\end{eqnarray}
Because the quark can interact directly with the virtual photon, the
evolution equation in Eq.~(\ref{Dq-NS-evo}) has an inhomogeneous term.  

By extending the simple ladder diagrams in Fig.~\ref{fig6} to 
general ladder diagrams \cite{CFP-DIS}, one can derive evolution
equations for the singlet quark-to-virtual-photon and
gluon-to-virtual-photon fragmentation functions \cite{QZ-VPFF}, 
\begin{eqnarray}
\mu_F^2 \frac{d}{d\mu_F^2} D_{c\rightarrow\gamma^* X}(z,\mu_F^2;Q^2)
&=& 
\left(\frac{\alpha_{em}}{2\pi}\right)
 \gamma_{c\rightarrow\gamma^*}(z,\mu_F^2,\alpha_s;Q^2)
\nonumber \\
&+&
\left(\frac{\alpha_{s}}{2\pi}\right)
\sum_d \int_z^1 \frac{dz'}{z'}
P_{c\rightarrow d}(\frac{z}{z'},\alpha_s)\, 
 D_{d\rightarrow\gamma^* X}(z',\mu_F^2;Q^2)\, ,
\label{RG-unpol}
\end{eqnarray}
where $c,d=q,\bar{q},g$.  In Eq.~(\ref{RG-unpol}), the evolution
kernels $P_{c\rightarrow d}$ are evaluated at a single hard scale,
$\mu_F$, and can be calculated perturbatively as a power series in 
$\alpha_s$.  QCD corrections to the QED quark-to-virtual-photon  
splitting function $\gamma_{c\rightarrow\gamma^*}$ can be evaluated 
in principle order-by-order in $\alpha_s$.

Calculating the lowest order quark-to-virtual-photon ladder diagram
in Fig.~\ref{fig6}, we obtain the leading order quark-to-virtual-photon
QED evolution kernel \cite{QZ-VPFF,BL-frag},
\begin{equation}
\gamma_{q\rightarrow \gamma^*}^{(0)}(z,k^2;Q^2) 
=
e_q^2 \Bigg[\frac{1+(1-z)^2}{z}
           -z\left(\frac{Q^2}{zk^2}\right) \Bigg]
      \theta(k^2-\frac{Q^2}{z})\, .
\label{Gq0-unpol}
\end{equation}
The $\theta$-function is a consequence of the mass threshold.
The gluon-to-virtual-photon evolution kernel vanishes at the same order 
\begin{equation}
\gamma_{g\rightarrow \gamma^*}^{(0)}(z,k^2;Q^2) 
= 0\, ,
\label{Gg0-unpol}
\end{equation}
because the gluon does not interact directly with the virtual photon.

It is important to note that if we work in the strict leading power
(or leading twist) approximation, we would drop {\it both} the power
corrections to the fragmentation functions as well as power
corrections to the evolution kernels of the fragmentation functions.
That is, we would neglect the $O(Q^2/k^2)$ term in
Eq.~(\ref{Gq0-unpol}),  
\begin{equation}
\gamma_{q\rightarrow \gamma^*}^{(LP-0)}(z,k^2) 
=\gamma_{q\rightarrow \gamma^*}^{(0)}(z,k^2;Q^2=0) 
=
e_q^2 \Bigg[\frac{1+(1-z)^2}{z}\Bigg]
      \theta(k^2-\frac{Q^2}{z})\, .
\label{Gq0-unpol-LP}
\end{equation}
The superscript ``LP'' represents the leading power
approximation.  With this strictly leading power quark-to-photon QED
evolution kernel, the evolution equation in Eq.~(\ref{RG-unpol})
for the parton-to-virtual-photon fragmentation functions is exactly
the same as that for the parton-to-real-photon fragmentation functions  
\cite{Owens-RMP}.  Under the strict leading power approximation, the 
only difference between virtual and real photon fragmentation
functions is the boundary condition for the evolution equations.  For
real photon fragmentation functions, a set of unknown
non-perturbative input fragmentation functions $D_{c\rightarrow
\gamma X}(z)$ is needed at a given scale $\mu_F^0$ ($\sim$ a few GeV)
\cite{Ph-frag}.  On the other hand, no non-perturbative input fragmentation 
functions are needed to solve the evolution equations in
Eq.~(\ref{RG-unpol}).  Instead, the mass threshold for production of a 
time-like virtual photon of invariant mass $Q$ imposes 
a natural boundary condition for all flavors $c$, 
\begin{equation}
D_{c\rightarrow \gamma^* X}(z,\mu_F^2\le Q^2/z;Q^2) = 0\, .
\end{equation}

The strict leading-power approximation might be too severe 
in the threshold region \cite{QZ-VPFF}.  Because its mass is non-zero,
the virtual photon can have both transverse and longitudinal
polarization modes.  The QED evolution kernel in
Eq.~(\ref{Gq0-unpol}) is a sum of evolution kernels for a quark to 
fragment into either transverse ($T$) or longitudinal ($L$)
polarization modes \cite{QZ-VPFF,BL-frag}, 
\begin{equation}
\gamma_{q\rightarrow \gamma^*}^{(0)}(z,k^2;Q^2) 
= 2\, \gamma_{q\rightarrow \gamma^*_T}^{(0)}(z,k^2;Q^2) 
+ \gamma_{q\rightarrow \gamma^*_L}^{(0)}(z,k^2;Q^2) , 
\label{Gq0-sum}
\end{equation}
with
\begin{eqnarray}
\gamma_{q\rightarrow \gamma^*_T}^{(0)}(z,k^2;Q^2) 
&=& e_q^2\, \frac{1}{2}\, 
\left(\frac{1+(1-z)^2}{z}\right)
 \left[1-\frac{Q^2}{z k^2}\right]
  \theta(k^2-\frac{Q^2}{z}) \, ;
\label{Gq0-T} \\
\gamma_{q\rightarrow \gamma^*_L}^{(0)}(z,k^2;Q^2) 
&=& e_q^2 
\left[2\left(\frac{1-z}{z}\right)\right]
       \left(\frac{Q^2}{z k^2}\right)
  \theta(k^2-\frac{Q^2}{z})\, .
\label{Gq0-L}
\end{eqnarray}
The factor $2$ in Eq.~(\ref{Gq0-sum}) represents the two transverse
polarization states of the virtual photon.  Under the strict leading
power approximation, Eq.~(\ref{Gq0-L}) vanishes.  In this strict 
limit, only transversely polarized virtual photons are produced through the
fragmentation processes, and there are no logarithmic contributions to the
production of longitudinally polarized virtual photons.
Furthermore, without the $O(Q^2/k^2)$ term in Eq.~(\ref{Gq0-T}), the
evolution kernel $\gamma_{q\rightarrow \gamma^*_T}^{(0)}(z,k^2;Q^2)$
gives the wrong threshold behavior.  Instead of being zero at the threshold
when $k^2=Q^2/z$, the leading power evolution kernel 
$\gamma_{q\rightarrow \gamma^*_T}^{(LP-0)}(z,k^2;Q^2)$ is finite and 
proportional to $1/z$.  It is large if $z$ is small. 

In general, there can be two types of power corrections to the leading
power virtual photon fragmentation functions. Power suppressed contributions 
$O(Q^2/\mu_F^2)$ to the fragmentation functions 
are one type (Type-I as defined in Ref.~\cite{QZ-DYresum}), and 
power corrections contributions to the evolution kernels (or the
slopes) of the fragmentation functions are the other type (known as Type-II). 
For example, consider the lowest order contribution to the
quark-to-virtual-photon fragmentation function from the evolution
kernel in Eq.~(\ref{Gq0-T}),  
\begin{eqnarray}
D_{q\rightarrow\gamma^*_T X}^{(0)}(z,\mu_F^2;Q^2) 
&\equiv &
\int_{Q^2/z}^{\mu_F^2}\, \frac{dk^2}{k^2}\, 
\left(\frac{\alpha_{em}}{2\pi}\right)
\gamma_{q\rightarrow \gamma^*_T}^{(0)}(z,k^2;Q^2)
\nonumber \\
& = &
e_q^2\, \left(\frac{\alpha_{em}}{2\pi}\right)\, \frac{1}{2}\,
\left(\frac{1+(1-z)^2}{z}\right)
 \left[\ln\left(\frac{z\mu_F^2}{Q^2}\right)
       -\left(1-\frac{Q^2}{z \mu_F^2}\right) \right]\, .
\label{Dq0-T}
\end{eqnarray}
The $(1-Q^2/z\mu_F^2)$ term results from the power suppressed
$Q^2/zk^2$ term in the evolution kernel, and it is clear that this
term is as important as the logarithmic term in the threshold
region.  The term proportional to 1 in this
$(1-Q^2/z\mu_F^2)$ combination is not power suppressed by
$O(Q^2/\mu_F^2)$.  In the threshold region, the virtual photon
fragmentation functions are dominated by the longitudinally polarized 
component particularly when $z$ is small \cite{QZ-VPFF}.  The Type-II 
power corrections are not necessarily small and could provide power 
{\em non-suppressed} contributions to physical observables \cite{QZ-DYresum}.

In the rest of our discussion, we keep the leading power
suppressed terms in the QED evolution kernels in Eq.~(\ref{Gq0-unpol})
when we calculate our parton-to-virtual-photon fragmentation functions.  
As shown in Ref.~\cite{QZ-VPFF}, the inhomogeneous QED evolution kernels
in Eq.~(\ref{RG-unpol}) dominate the scale
dependence of the fragmentation functions, and therefore, we neglect
the power corrections to the QCD evolution kernels $P_{c\rightarrow
d}$ in Eq.~(\ref{RG-unpol}).  With the inclusion of power
corrections in the evolution kernels, the resummation discussed here
is no longer a simple one-scale problem in QCD perturbation theory.
More detailed discussions of the virtual-photon fragmentation
functions can be found in Ref.~\cite{QZ-VPFF}.

In summary, the all-orders resummation of the large logarithmic
contributions to the low mass Drell-Yan cross section is equivalent
to a sum of all logarithmic contributions to the virtual photon
fragmentation functions, achieved by solution of the evolution
equations in Eq.~(\ref{RG-unpol}).  The evolution equations 
for the virtual photon fragmentation functions have the same
functional forms as those for real photon fragmentation functions
\cite{Owens-RMP}. However, the differences in both boundary conditions and 
the inhomogeneous terms due to the non-vanishing of $Q^2$ lead to many 
differences between the real and virtual photon fragmentation
functions.  One major difference is that the virtual photon
fragmentation functions are purely perturbative.  In addition, the
virtual photon fragmentation functions provide the resummed
contributions to the production of a longitudinally polarized 
virtual photon \cite{QZ-VPFF}. 

%========================================================================
\subsection{Calculation of the Partonic Hard Parts}
\label{sec3c}

To determine the fragmentation contributions to the Drell-Yan
cross section, 
\begin{eqnarray}
\frac{d\sigma_{AB\rightarrow \gamma^*(Q) X}^{(F)}}{dQ_T^2\,dy}
&=&
\sum_{a,b}\int dx_1 \phi_{a/A}(x_1,\mu) 
           \int dx_2 \phi_{b/B}(x_2,\mu)\,
\nonumber \\
&\times &
\sum_{c} \int \frac{dz}{z^2}\, \left[
  \frac{d\hat{\sigma}_{ab\rightarrow c X}^{(F)}}{dp_{c_T}^2\,dy}
  \left(x_1,x_2,p_c=\hat{Q}/z;\mu_F\right)  \right]
  D_{c\rightarrow \gamma^* X}(z,\mu_F^2;Q^2)\, ,
\label{DY-F-fac}
\end{eqnarray}
we must evaluate the partonic hard parts 
$d\hat{\sigma}_{ab\rightarrow c X}^{(F)}/dp_{c_T}^2 dy$ in
Eq.~(\ref{DY-F-fac}).  Although these are calculable 
perturbatively, exact expressions depend on how the
parton distributions and fragmentation functions are defined (or 
the choice of factorization scheme).  In this subsection, we provide a
self-consistent procedure for the calculation of the partonic hard parts.

We separate the procedure into four steps: (1) instead of considering 
the hadronic process: $A+B\rightarrow \gamma^*(Q) X$, we 
apply the factored formula in
Eq.~(\ref{DY-F-fac}) to a partonic process: $a'+b'\rightarrow c' X$ with
an on-shell final-state parton $c'$; (2) we expand both sides of the
factored formula for the partonic process order-by-order in
$\alpha_s$; (3) we calculate the partonic cross section 
$d\sigma_{a'b'\rightarrow c' X}/dp_{c'_T}^2 dy$, and parton-to-parton
distributions and fragmentation functions order-by-order in
$\alpha_s$; and (4) we extract the short-distance partonic hard parts
$d\hat{\sigma}_{ab\rightarrow c X}/dp_{c_T}^2 dy$ by comparing both
sides of the perturbatively expanded factored formula at the same
order of $\alpha_s$.

By applying the fragmentation expression in
Eq.~(\ref{DY-F-fac}) to the partonic processes: $a'+b'\rightarrow c' X$,
we obtain the following schematic formula
\begin{equation}
\sigma_{a'b'\rightarrow c'} = \sum_{a,b,c} 
\phi_{a/a'} \otimes \phi_{b/b'} \otimes 
\hat{\sigma}_{ab\rightarrow c}^{(F)} \otimes D_{c\rightarrow c'} .
\label{DY-F-pfac}
\end{equation}
Symbol $\otimes$ represents the convolutions over momentum fractions
$x_1$, $x_2$, and $z$ in Eq.~(\ref{DY-F-fac}).  (In Eq.~(\ref{DY-F-pfac}) 
and in the rest of the equations of this subsection, we omit the inclusive 
symbol $X$ for reasons of notational simplicity.)

To produce a parton with large transverse momentum, a 
$2\rightarrow 2$ partonic process is required at the parton level, at 
minimum of the order $O(\alpha_s^2)$.  Expanding both sides of the factored 
formula Eq.~(\ref{DY-F-pfac}) order-by-order in $\alpha_s$, we define 
the following perturbative expansions, 
\begin{eqnarray}
\sigma_{a'b'\rightarrow c'}
&\equiv & 
\sum_{n=2}\, \sigma^{(n)}_{a'b'\rightarrow c'}
\left(\frac{\alpha_s(\mu)}{2\pi}\right)^n\, ,
\label{S-ab-n} \\
\hat{\sigma}_{ab\rightarrow c}^{(F)}
&\equiv & 
\sum_{n=2}\, H^{(n)}_{ab\rightarrow c}
\left(\frac{\alpha_s(\mu)}{2\pi}\right)^n\, ,
\label{H-ab-n} \\
\phi_{a/a'}
&\equiv &
\sum_{n=0}\, \phi_{a/a'}^{(n)}
\left(\frac{\alpha_s(\mu)}{2\pi}\right)^n\, ,
\label{f-a-n} \\
D_{c\rightarrow c'}
&\equiv &
\sum_{n=0}\, D_{c\rightarrow c'}^{(n)}
\left(\frac{\alpha_s(\mu)}{2\pi}\right)^n\, .
\label{D-c-n}
\end{eqnarray}
We substitute these four perturbative expansions into
Eq.~(\ref{DY-F-pfac}) and obtain at $O(\alpha_s^2)$, 
\begin{equation}
\sigma_{a'b'\rightarrow c'}^{(2)} = \sum_{a,b,c}
\phi_{a/a'}^{(0)} \otimes \phi_{b/b'}^{(0)} \otimes 
H_{ab\rightarrow c}^{(2)} \otimes D_{c\rightarrow c'}^{(0)}\, .
\label{DY-F-pfac-2}
\end{equation}
Since the zeroth order parton distributions and fragmentation
functions are $\delta$-functions, Eq.~(\ref{DY-F-pfac-2})
yields $H_{ab\rightarrow c}^{(2)} = \sigma_{ab\rightarrow c}^{(2)}$ or 
\begin{equation}
\frac{d\hat{\sigma}_{ab\rightarrow c}^{(F-LO)}}{dp_{c_T}^2\,dy}
= \frac{d\sigma_{ab\rightarrow c}^{(LO)}}{dp_{c_T}^2\,dy}
\label{H-ab-LO}
\end{equation}
at leading order.  

Expanding Eq.~(\ref{DY-F-pfac}) to NLO, we write 
\begin{eqnarray}
\sigma_{a'b'\rightarrow c'}^{(3)} 
&= & \sum_{a,b,c}
\phi_{a/a'}^{(0)} \otimes \phi_{b/b'}^{(0)} \otimes 
H_{ab\rightarrow c}^{(3)} \otimes D_{c\rightarrow c'}^{(0)}
\nonumber \\
&+& \sum_{a,b,c}
\phi_{a/a'}^{(1)} \otimes \phi_{b/b'}^{(0)} \otimes 
H_{ab\rightarrow c}^{(2)} \otimes D_{c\rightarrow c'}^{(0)}
\nonumber \\
&+& \sum_{a,b,c}
\phi_{a/a'}^{(0)} \otimes \phi_{b/b'}^{(1)} \otimes 
H_{ab\rightarrow c}^{(2)} \otimes D_{c\rightarrow c'}^{(0)}
\nonumber \\
&+& \sum_{a,b,c}
\phi_{a/a'}^{(0)} \otimes \phi_{b/b'}^{(0)} \otimes 
H_{ab\rightarrow c}^{(2)} \otimes D_{c\rightarrow c'}^{(1)}\, .
\label{DY-F-pfac-3}
\end{eqnarray}
Using the zero-th order parton distributions and fragmentation
functions, for $d\hat{\sigma}_{ab\rightarrow c}^{(F-NLO)}/dp_{c_T}^2\,dy$, 
we obtain 
\begin{eqnarray}
H_{ab\rightarrow c}^{(3)} 
&=& \sigma_{ab\rightarrow c}^{(3)} 
- \sum_{a'} \phi_{a'/a}^{(1)} \otimes H_{a'b\rightarrow c}^{(2)}
\nonumber \\
&-& 
  \sum_{b'} \phi_{b'/b}^{(1)} \otimes H_{ab'\rightarrow c}^{(2)} 
- \sum_{c'} H_{ab\rightarrow c'}^{(2)} 
            \otimes D_{c'\rightarrow c}^{(1)} .
\label{H-ab-NLO}
\end{eqnarray}
In Eq.~(\ref{H-ab-NLO}), $H_{ab\rightarrow c}^{(2)}$ is the leading
order contribution calculated in Eq.~(\ref{H-ab-LO}).  
The partonic cross section $\sigma_{ab\rightarrow c}^{(3)}$ and all
parton-level parton distributions and fragmentation functions are 
perturbatively calculable with proper regulators. The
subtraction terms in Eq.~(\ref{H-ab-NLO}) remove the collinear
singularities associated with the massless partons.  
Following the same procedure, we can derive the short-distance
partonic hard parts for the fragmentation contributions in
Eq.~(\ref{DY-F-fac}) at all orders in $\alpha_s$.  

Equation~(\ref{H-ab-NLO}) shows that beyond the leading order, the 
exact expressions for the perturbatively calculated short-distance hard 
parts depend on the definitions of parton-level parton distributions and 
fragmentation functions.  The partonic hard
parts are fixed uniquely once we fix the parton-level parton
distributions and fragmentation functions.  In order to use 
available conventional parton distributions, we have little 
choice other than to select the parton-level parton distributions in
either the $\overline{MS}$ or $DIS$ scheme~\cite{CTEQ-bk}.  

In the partonic cross section $\sigma^{(n)}_{ab\rightarrow c}$, the 
parton momentum $p_c$ is assumed to be massless, $p_c^2=0$, and, 
therefore, a final-state collinear singularity arises. 
Within the usual QCD factorization framework, there is freedom to
choose any factorization scheme to remove the final-state collinear
singularities of the partonic cross section and absorb all possible
finite differences into the non-perturbative fragmentation functions.
Different choices for the factorization scheme lead to finite
differences between the extracted non-perturbative fragmentation
functions.  However, owing to the non-zero invariant mass of the
virtual photon, the parton-to-virtual-photon fragmentation functions
do not have final-state collinear singularities.  They are completely 
perturbative.  Therefore, the parton-to-virtual-photon fragmentation 
functions cannot uniquely fix the definition of the parton-to-parton 
fragmentation functions.  As a consequence of the difference in the 
invariant masses of the parton and the virtual photon, an extra constraint 
has to be introduced to specify the parton-to-parton fragmentation functions 
$D^{(n)}_{c'\rightarrow c}$.  We can choose a scheme for defining the
parton-level fragmentation functions so as to remove the final-state
collinear singularities in the partonic cross sections  
$\sigma_{ab\rightarrow c}^{(n)}$, with $n\ge 3$.  The finite
differences between schemes cannot be completely absorbed into the
{\it perturbative} parton-to-virtual-photon fragmentation functions.

Finite differences associated with the choice of scheme for the
parton-level fragmentation functions correspond to non-logarithmic
contributions to the Drell-Yan cross section.  For 
the logarithmic contributions to the Drell-Yan fragmentation functions, 
we can choose parton-level fragmentation functions 
in the $\overline{MS}$ scheme or in any other scheme when calculating 
the partonic short-distance hard
parts in Eq.~(\ref{DY-F-pfac-3}).  The perturbatively calculated 
partonic hard parts $d\hat{\sigma}_{ab\rightarrow c}^{(F)}/dp_{c_T}^2
dy$ will be the same as those for inclusive single hadron (or prompt
photon) production if the same factorization scheme is used.
The non-logarithmic differences caused by the different
choices of the factorization schemes can be absorbed
into the {\em direct} contributions to the Drell-Yan cross section,
defined and discussed in next section.

%%%%%%%%%%%%%%%%%%%%%%%%%%%%%%%%%%%%%%%%%%%%%%%%%%%%%%%%%%%%%%%%%%%%%%%%%
\section{Drell-Yan Cross Section with Resummed Fragmentation
Contributions} 
\label{sec4}

In this section, we derive our modified factorization formula for the 
Drell-Yan cross section including resummation of the logarithmic
contributions.

Particularly at leading order, but also at all higher orders in perturbation 
theory, there are significant contributions to the cross section that 
are not included in the fragmentation terms.  It is essential that 
the full factorization formula include all non-logarithmic contributions 
order-by-order in $\alpha_s$, while retaining the all orders resummation 
of the logarithmic contributions.  

We use the word {\em direct} to designate the non-logarithmic 
contributions to the Drell-Yan cross section .  This use of the term  
is the same as its use in prompt photon production.  At leading order the 
direct terms in the Drell Yan cross section are those 
supplied by the short-distance $O(1/Q_T)$ 2-parton to 2-parton Compton and 
annihilation subprocesses.  However, there is an additional component at 
higher orders.  The direct contribution must also absorb the finite non-logarithmic 
differences in the partonic hard parts $H_{ab\rightarrow c X}^{(m)}$ associated 
with the ambiguity in definition of the parton-level fragmentation functions, as 
discussed in the last section.  This second component accounts for the 
non-logarithmic terms over the distance interval $O(1/Q)$ to $O(1/Q_T)$.  The 
logarithmic terms are included in the fragmentation functions.  The 
physics of our direct term is very similar to that represented by the $Y$ 
term in the CSS formalism~\cite{CSS-DYresum} for resummation at small $Q_T$.   

To be precise, we {\em define} the direct contribution as the difference
\begin{eqnarray}
\frac{d\sigma_{AB\rightarrow \gamma^*(Q) X}^{(Dir)}}{dQ_T^2\,dy}
&\equiv &
\frac{d\sigma_{AB\rightarrow \gamma^*(Q) X}}{dQ_T^2\,dy}
- 
\frac{d\sigma_{AB\rightarrow \gamma^*(Q) X}^{(F)}}{dQ_T^2\,dy}
\nonumber \\
& = &
\sum_{a,b}\int dx_1 \phi_{a/A}(x_1,\mu) 
           \int dx_2 \phi_{b/B}(x_2,\mu)\,
\left[
 \frac{d\hat{\sigma}_{ab\rightarrow \gamma^*(Q) X}^{(Dir)}}{dQ_T^2\,dy}  
\right] .
\label{DY-dir}
\end{eqnarray}
The parton level direct term is obtained from the factored formulas 
in Eqs.~(\ref{Vph-fac}) and (\ref{DY-F-fac}), 
\begin{eqnarray}
\frac{d\hat{\sigma}^{(Dir)}_{ab\rightarrow \gamma^*(Q) X}}{dQ_T^2\,dy}
 \left(x_1,x_2,Q,Q_T,y;\mu,\mu_F\right)
&\equiv & 
\frac{d\hat{\sigma}_{ab\rightarrow \gamma^*(Q) X}}{dQ_T^2\,dy}
 \left(x_1,x_2,Q,Q_T,y;\mu\right) 
\nonumber \\
&-&
\frac{d\hat{\sigma}^{(F)}_{ab\rightarrow \gamma^*(Q) X}}{dQ_T^2\,dy}
 \left(x_1,x_2,Q,Q_T,y;\mu,\mu_F\right)\, ,
\label{DY-pdir}
\end{eqnarray}
with $d\hat{\sigma}^{(F)}_{ab\rightarrow \gamma^*(Q) X}/dQ_T^2\,dy$
given in Eq.~(\ref{DY-F}) and 
$d\hat{\sigma}_{ab\rightarrow \gamma^*(Q) X}/dQ_T^2\,dy$ calculated in
conventional~fixed-order~QCD~perturbation~theory.~Since 
$d\hat{\sigma}_{ab\rightarrow \gamma^*(Q) X}/dQ_T^2\,dy$
and $d\hat{\sigma}_{ab\rightarrow \gamma^*(Q) X}^{(F)}/dQ_T^2\,dy$ are 
calculable perturbatively, the direct contributions,
$d\hat{\sigma}^{(Dir)}_{ab\rightarrow \gamma^*(Q) X}/dQ_T^2\,dy$ in
Eq.~(\ref{DY-pdir}), should also be calculable perturbatively.  They 
have the perturbative expansion  
\begin{equation}
\frac{d\hat{\sigma}^{(Dir)}_{ab\rightarrow \gamma^*(Q) X}}{dQ_{T}^2\,dy}
=\sum_{n=1} 
Y^{(n)}_{ab\rightarrow \gamma^*(Q) X}(x_1,x_2,Q,Q_T,y;\mu_F,\mu)
\left(\frac{\alpha_s(\mu)}{2\pi}\right)^n\, ,
\label{Y-perp}
\end{equation}
where $n\ge 1$ because the leading order contributions to
Drell-Yan cross section are of order $O(\alpha_{em}\alpha_s)$.
Since the fragmentation contributions
$d\hat{\sigma}_{ab\rightarrow \gamma^*(Q) X}^{(F)}/dQ_T^2\,dy$ and 
$d\hat{\sigma}_{ab\rightarrow \gamma^*(Q) X}/dQ_T^2\,dy$ share the same   
large logarithmic terms, $\ln^m(Q_T^2/Q^2)$, the direct contributions 
should be free of large logarithms order-by-order in $\alpha_s$.  Therefore,
the coefficient functions $Y^{(n)}_{ab\rightarrow \gamma^*(Q) X}$ in
Eq.~(\ref{Y-perp}) are evaluated at a single hard scale
($\sim Q_T$), and the perturbative expansion for the direct
contribution in Eq.~(\ref{Y-perp}) is expected to be well-behaved
perturbatively.  We substitute Eqs.~(\ref{Vph-Hab}),
(\ref{H-ab-n}), (\ref{D-c-n}), and (\ref{Y-perp}) into
Eq.~(\ref{DY-pdir}) and obtain 
\begin{equation}
Y^{(n)}_{ab\rightarrow \gamma^*(Q) X}
=H^{(n)}_{ab\rightarrow \gamma^*(Q) X}
-\sum_{m=2}^{n} \left[\sum_{c}
  H^{(m)}_{ab\rightarrow c X} \otimes 
  D_{c\rightarrow\gamma^*(Q) X}^{(n-m)}\right]\, .
\label{Y-ab}
\end{equation}
Functions $D_{c\rightarrow\gamma^*(Q) X}^{(n-m)}$ are the coefficient
functions for the perturbatively calculated parton-to-virtual-photon
fragmentation functions \cite{QZ-VPFF}. 
The subtraction term in Eq.~(\ref{Y-ab}) removes the
logarithmic contributions included in the coefficient
functions $H_{ab\rightarrow \gamma^*(Q) X}$ that are calculated in  
conventional fixed-order perturbation theory. As remarked above, 
the direct contributions defined in Eq.~(\ref{Y-ab}) absorb the finite 
non-logarithmic differences in the partonic hard parts 
$H_{ab\rightarrow c X}^{(m)}$.  If $H^{(m)}_{ab\rightarrow c X}$ is larger 
in one scheme and more is included in the fragmentation contributions,
then more will be subtracted from the direct contributions.  Any
non-compensated differences will be higher order corrections in
$\alpha_s$. 
 
We substitute Eq.~(\ref{DY-pdir}) into Eq.~(\ref{DY-dir}) and derive a
modified factorization formula for the Drell-Yan cross
section at $Q_T\ge Q$:
\begin{eqnarray}
\frac{d\sigma_{AB\rightarrow \gamma^*(Q) X}}{dQ_T^2\,dy}
&=&\sum_{a,b} \int dx_1 \phi_{a/A}(x_1,\mu) 
              \int dx_2 \phi_{b/B}(x_2,\mu)
\nonumber\\
&\ & \times \Bigg\{ 
 \frac{d\hat{\sigma}_{ab\rightarrow \gamma^*(Q) X}^{(Dir)}}{dQ_T^2\,dy}
        (x_1,x_2,Q,Q_T,y;\mu_F,\mu)
\label{DY-mfac} \\
&\ & {\hskip 0.2in} +
\sum_c \int\frac{dz}{z^2} \left[
 \frac{d\hat{\sigma}_{ab\rightarrow c X}^{(F)}}{dp_{c_T}^2\,dy}
        (x_1,x_2,p_c=\frac{\hat{Q}}{z};\mu_F,\mu) \right]
 D_{c\rightarrow\gamma^*(Q) X}(z,\mu_F^2;Q^2)
 \Bigg\} \, .
\nonumber 
\end{eqnarray}
Both perturbatively calculated short-distance hard parts,
$d\hat{\sigma}_{ab\rightarrow c X}^{(F)}/dp_{c_T}^2\,dy$ and 
$d\hat{\sigma}_{ab\rightarrow \gamma^*(Q) X}^{(Dir)}/dQ_T^2\,dy$, are
free of large logarithms.  All potentially large $\ln^m(Q_T^2/Q^2)$
terms in the region $Q_T^2\gg Q^2$ are resummed into the virtual photon 
fragmentation functions $D_{c\rightarrow\gamma^*(Q) X}$. The direct 
contribution (first term on 
the right-hand side of Eq.~(\ref{DY-mfac})) represents the production 
of lepton-pairs at the distance scale of the hard collisions ($\sim 1/Q_T$).  
The fragmentation contribution stands for the sum of all 
leading logarithmic contributions from a distance scale at $1/Q_T$ to
$1/Q$.  

We note here, in passing, that unlike the logarithmic terms in the region of 
small $Q_T$, the logarithmic terms at large $Q_T$ are typical DGLAP logarithms 
associated with collinear contributions.  At low $Q_T$, logarithmic contributions 
arise from the collinear {\em and} the infrared regions.  Correspondingly, in 
resummation formalisms relevant at small $Q_T$, there are leading log terms 
(including both collinear and infrared logs at all orders in $\alpha_s$), and  
next-to-leading log terms (including only one of the two types of logs), and 
``next-to-next-to-leading" log terms, and so forth.  In our case, we do not have 
the type of ``next-to-leading" logs that appear in resummation procedures at small 
$Q_T$.  Anything left-over in the large $Q_T$ region, after resummation of the 
leading logs of DGLAP type, is included in the direct terms.

Since we have in mind applications at large $Q_T$ but small $Q$, it is 
important to consider the possible role of higher-twist contributions proportional 
to inverse-powers of $Q$.   Like power corrections from target mass effects, 
power corrections in our case should appear in the form of 
$m^2/({\rm energy \, exchange})^2$, with $m \sim Q$.  Since $Q_T^2$ sets the hard 
scale, 
and $Q^2$ acts as a mass threshold, the only dimensionless ratios we expect to see 
are $Q^2/Q_T^2$ and $Q^2/\mu^2$, but no $1/Q^2$ term. There can be three kinds of 
power corrections: power corrections to the partonic hard parts, suppressed by 
$Q^2/Q_T^2$; power corrections to the fragmentation contributions, suppressed by 
$Q^2/\mu_F^2 \sim Q^2/Q_T^2$; and power corrections to the evolution kernels of 
the fragmentation functions, proportional to $Q^2/\mu^2$.  Since $\mu^2$
runs from $Q^2$ to $\mu_F^2$, the third type could be significant, 
$Q^2/\mu_{min}^2 \sim O(1)$.  Although this third variety is potentially important, 
the terms should be much smaller than the leading logarithmic 
contributions as long as $\ln(Q_T^2/Q^2) \gg 1$.   

Our modified factorization formula for the Drell-Yan cross section in
Eq.~(\ref{DY-mfac}) is very similar to that for prompt real photon
production in Eq.~(\ref{Ph-fac}).  However, the $Q^2$-dependence
in the direct production term and differences in the fragmentation
functions distinguish the Drell-Yan virtual photon production
from prompt real photon production. 

The key difference between our modified factorization formula in
Eq.~(\ref{DY-mfac}) and the conventional factorization formula in
Eq.~(\ref{Vph-fac}) resides in the way the logarithmic contributions
from final-state parton splitting are handled.  Instead of one 
perturbative series in powers of $\alpha_s$ in the conventional approach, 
we have two perturbative expansions in our
modified factorization formula: one for the direct and one for the
fragmentation contribution.  All coefficient functions in the new
perturbative expansions are free of large logarithms.  The large
logarithms in the conventional perturbative expansion are resummed
into the fragmentation functions.  In the rest of this section, we
demonstrate the difference by comparing the two factorization formulas 
explicitly order-by-order in powers of $\alpha_s$.

According to the conventional factorization approach, the lowest order
partonic contributions to virtual photon production are provided 
by the quark-gluon Compton and quark-antiquark annihilation
diagrams at order $\alpha_{em}\alpha_s$:
\begin{equation}
\frac{d\sigma_{AB\rightarrow \gamma^*(Q) X}^{(C-LO)}}{dQ_T^2\,dy}
=
\sum_{a,b} \int dx_1 \phi_{a/A}(x_1,\mu) 
                \int dx_2 \phi_{b/B}(x_2,\mu) \left[
     \frac{d\hat{\sigma}_{ab\rightarrow \gamma^*(Q) X}^{(C-LO)}}
          {dQ_T^2\,dy}
        (x_1,x_2,Q,Q_T,y;\mu) \right] \, .
\label{DY-c-LO}
\end{equation}
The superscript $(C-LO)$ stands for the leading order
conventional perturbative calculation. 

In our modified factorization approach, 
Eq.~(\ref{DY-mfac}), there are two perturbative expansions corresponding
to the direct and fragmentation contributions, respectively.
Each perturbation series has its own leading order contributions.
Because the subtraction term in Eq.~(\ref{Y-ab}) starts at
$O(\alpha_s^2)$, the LO direct contribution is the same as the LO
conventional contribution in Eq.~(\ref{DY-c-LO}),
\begin{equation}
\frac{d\sigma_{AB\rightarrow \gamma^*(Q) X}^{(Dir-LO)}}{dQ_T^2\,dy}
=
\sum_{a,b} \int dx_1 \phi_{a/A}(x_1,\mu) 
                \int dx_2 \phi_{b/B}(x_2,\mu) \left[
     \frac{d\hat{\sigma}_{ab\rightarrow \gamma^*(Q) X}^{(Dir-LO)}}
          {dQ_T^2\,dy}
        (x_1,x_2,Q,Q_T,y;\mu) \right] \, .
\label{DY-dir-LO}
\end{equation}
The LO partonic hard part  
$d\hat{\sigma}_{ab\rightarrow\gamma^*(Q) X}^{(Dir-LO)}/dQ_T^2 dy$ 
equals $d\hat{\sigma}_{ab\rightarrow\gamma^*(Q) X}^{(C-LO)}/dQ_T^2 dy$ in
Eq.~(\ref{DY-c-LO}). 

The virtual photon fragmentation functions are of order 
$\alpha_{em}$ \cite{QZ-VPFF}, and the partonic hard parts, 
$d\hat{\sigma}_{ab\rightarrow c X}^{(F)}/dp_{c_T}^2\,dy$
in Eq.~(\ref{DY-mfac}) start at order $\alpha_s^2$.  The LO 
fragmentation contributions to the Drell-Yan cross section would seem
therefore to be of order $O(\alpha_{em}\alpha_s^2)$, one power
of $\alpha_s$ higher than the LO direct contributions in
Eq.~(\ref{DY-dir-LO}).  However, the large logarithms
from the virtual photon fragmentation functions are proportional to 
$\ln(\mu_F^2) \propto 1/\alpha_s(\mu_F)$, meaning that the LO fragmentation 
contributions to the Drell-Yan cross section can be viewed as 
effectively $O(\alpha_{em}\alpha_s)$ terms, of the same order as the 
LO direct contribution: 
\begin{eqnarray}
\frac{d\sigma_{AB\rightarrow \gamma^*(Q) X}^{(F-LO)}}
     {dQ_T^2\,dy}
&=&\sum_{a,b} \int dx_1 \phi_{a/A}(x_1,\mu) 
                \int dx_2 \phi_{b/B}(x_2,\mu)
\nonumber\\
&\ & \times 
     \sum_c \int \frac{dz}{z^2}\, 
     \frac{d\hat{\sigma}_{ab\rightarrow c X}^{(F-LO)}}
          {dp_{c_T}^2\,dy}
          (x_1,x_2,p_c=\frac{\hat{Q}}{z};\mu_F,\mu)\,
          D_{c\rightarrow\gamma^*(Q) X}(z,\mu_F^2;Q^2)\, .
\label{DY-F-LO}
\end{eqnarray}
In our modified factorization formalism, the LO contribution to the 
Drell-Yan cross section at large $Q_T$ is equal to the sum of the LO terms 
of the two perturbative expansions:
\begin{equation}
\frac{d\sigma_{AB\rightarrow \gamma^*(Q) X}^{(LO)}}{dQ_T^2\,dy}
=
\frac{d\sigma_{AB\rightarrow \gamma^*(Q) X}^{(Dir-LO)}}{dQ_T^2\,dy}
+ 
\frac{d\sigma_{AB\rightarrow \gamma^*(Q) X}^{(F-LO)}}{dQ_T^2\,dy} .
\label{DY-LO}
\end{equation}
The LO direct and fragmentation contributions are found in
Eq.~(\ref{DY-dir-LO}) and Eq.~(\ref{DY-F-LO}), respectively.

At NLO, $\alpha_{em}\alpha_s^2$, in the conventional fixed-order approach, 
the Drell-Yan cross section has the same form as that 
in Eq.~(\ref{DY-c-LO}) but with superscripts $(C-LO)$ replaced by $(C-NLO)$.

According to our modified factorization formalism, the NLO term should be 
the sum of the NLO terms in {\it both} the direct and 
fragmentation contributions in Eq.~(\ref{DY-mfac}).  The NLO direct 
contribution is 
\begin{eqnarray}
\frac{d\sigma_{AB\rightarrow \gamma^*(Q) X}^{(Dir-NLO)}}
     {dQ_T^2\,dy}
&=&\sum_{a,b} \int dx_1 \phi_{a/A}(x_1,\mu) 
                \int dx_2 \phi_{b/B}(x_2,\mu)
\nonumber\\
&\ & \times \Bigg[
     \frac{d\hat{\sigma}_{ab\rightarrow \gamma^*(Q) X}^{(C-NLO)}}
          {dQ_T^2\,dy}
          (x_1,x_2,Q,Q_T,y;\mu)
\label{DY-dir-NLO} \\
&\ & {\hskip 0.1in}
   - \sum_c \int \frac{dz}{z^2}\, 
     \frac{d\hat{\sigma}_{ab\rightarrow c X}^{(F-LO)}}
          {dp_{c_T}^2\,dy}
          (x_1,x_2,p_c=\frac{\hat{Q}}{z};\mu_F,\mu)\,
          D_{c\rightarrow\gamma^*(Q) X}^{(0)}(z,\mu_F^2;Q^2) \Bigg]\, .
\nonumber
\end{eqnarray}
The subtraction term is a consequence of the definition of the
direct contribution in Eq.~(\ref{Y-ab}).  The subtraction term is
also necessary to remove the logarithmic terms from the
conventional NLO expression in Eq.~(\ref{DY-dir-NLO}), to avoid double 
counting. These
logarithmic contributions have been included in the LO fragmentation
contribution to the Drell-Yan cross section in Eq.~(\ref{DY-LO}).   

We would need NLO fragmentation contributions to complete the calculation of 
the NLO contribution,
\begin{eqnarray}
\frac{d\sigma_{AB\rightarrow \gamma^*(Q) X}^{(F-NLO)}}
     {dQ_T^2\,dy}
&=&\sum_{a,b} \int dx_1 \phi_{a/A}(x_1,\mu) 
                \int dx_2 \phi_{b/B}(x_2,\mu)
\nonumber\\
&\ & \times 
     \sum_c \int \frac{dz}{z^2}\, 
     \frac{d\hat{\sigma}_{ab\rightarrow c X}^{(F-NLO)}}
          {dp_{c_T}^2\,dy}
          (x_1,x_2,p_c=\frac{\hat{Q}}{z};\mu_F,\mu)\,
          D_{c\rightarrow\gamma^*(Q) X}(z,\mu_F^2;Q^2)\, .
\label{DY-F-NLO}
\end{eqnarray}
The virtual-photon fragmentation functions include the all orders 
resummation of the logarithmic terms, and they are the same as those for 
the LO contribution in Eq.~(\ref{DY-F-LO}).  The $O(\alpha_S^3)$ partonic hard 
parts $d\hat{\sigma}_{ab\rightarrow c X}^{(F-NLO)}/dp_{c_T}^2\,dy$ are defined 
in Sec.~\ref{sec3c} and will be presented elsewhere.

The NLO contribution to the Drell-Yan cross section at large
transverse momentum is  
\begin{equation}
\frac{d\sigma_{AB\rightarrow \gamma^*(Q) X}^{(NLO)}}
     {dQ_T^2\,dy}
\equiv
\frac{d\sigma_{AB\rightarrow \gamma^*(Q) X}^{(Dir-NLO)}}
     {dQ_T^2\,dy}
+
\frac{d\sigma_{AB\rightarrow \gamma^*(Q) X}^{(F-NLO)}}
     {dQ_T^2\,dy}\, ,
\label{DY-NLO}
\end{equation}
where the NLO direct and fragmentation contributions are found in
Eqs.~(\ref{DY-dir-NLO}) and (\ref{DY-F-NLO}), respectively.

In our modified factorization formalism, the LO and NLO contributions are 
different from those in the conventional formalism because the virtual 
photon fragmentation functions include all orders of the large leading 
logarithmic contributions.  The difference is better seen if we rewrite 
our LO and NLO results in the following form, 
\begin{eqnarray}
\frac{d\sigma_{AB\rightarrow \gamma^*(Q) X}}
     {dQ_T^2\,dy}
&\equiv &
\frac{d\sigma_{AB\rightarrow \gamma^*(Q) X}^{(LO)}}
     {dQ_T^2\,dy}
+
\frac{d\sigma_{AB\rightarrow \gamma^*(Q) X}^{(NLO)}}
     {dQ_T^2\,dy}
\label{DY-m-LO-NLO} \\
&=&\sum_{a,b} \int dx_1 \phi_{a/A}(x_1,\mu) 
                \int dx_2 \phi_{b/B}(x_2,\mu)
\nonumber\\
&\ & \times \left\{
     \frac{d\hat{\sigma}_{ab\rightarrow \gamma^*(Q) X}^{(C-LO)}}
          {dQ_T^2\,dy}
          (x_1,x_2,Q,Q_T,y;\mu)
   + \frac{d\hat{\sigma}_{ab\rightarrow \gamma^*(Q) X}^{(C-NLO)}}
          {dQ_T^2\,dy}
          (x_1,x_2,Q,Q_T,y;\mu) \right\}
\nonumber\\
&+&\sum_{a,b} \int dx_1 \phi_{a/A}(x_1,\mu) 
                \int dx_2 \phi_{b/B}(x_2,\mu)
\nonumber\\
&\ & \times \Bigg\{
   \sum_c \int \frac{dz}{z^2}\, 
     \frac{d\hat{\sigma}_{ab\rightarrow c X}^{(F-LO)}}
          {dp_{c_T}^2\,dy}
          (x_1,x_2,p_c=\frac{\hat{Q}}{z};\mu_F,\mu)
\nonumber\\
&\ & {\hskip 0.4in} \times
     \left[D_{c\rightarrow\gamma^*(Q) X}(z,\mu_F^2;Q^2)
          -D_{c\rightarrow\gamma^*(Q) X}^{(0)}(z,\mu_F^2;Q^2) \right]
\label{DY-m-LO-NLO-2} \\
&\ & {\hskip 0.1in}
  +\sum_c \int \frac{dz}{z^2}\, 
     \frac{d\hat{\sigma}_{ab\rightarrow c X}^{(F-NLO)}}
          {dp_{c_T}^2\,dy}
          (x_1,x_2,p_c=\frac{\hat{Q}}{z};\mu_F,\mu)
     D_{c\rightarrow\gamma^*(Q) X}(z,\mu_F^2;Q^2) 
  \Bigg\} .
\nonumber
\end{eqnarray}
On the right-hand-side of Eq.~(\ref{DY-m-LO-NLO-2}), the first term 
the prediction of the conventional fixed-order
factorization formalism in Eq.~(\ref{Vph-fac}); the second term
is the difference between our modified 
factorization formula and the conventional fixed-order factorization
formula up to NLO.  

To conclude this section, we emphasize that our modified factorization
formalism in Eq.~(\ref{DY-mfac}) effectively reorganizes the
{\it single} perturbative expansion of conventional QCD
factorization, $d\hat{\sigma}_{ab\rightarrow\gamma^*(Q) X}/dQ_T^2 dy$,
in Eq.~(\ref{Vph-fac}), into {\it two} perturbative expansions,
$d\hat{\sigma}_{ab\rightarrow\gamma^*(Q) X}^{(Dir)}/dQ_T^2 dy$ and 
$d\hat{\sigma}_{ab\rightarrow c X}^{(F)}/dp_{c_T}^2 dy$, plus the
perturbatively calculated evolution kernels.  The main advantage of
this reorganization is that the new perturbative expansions are 
evaluated at a single hard scale and are free of large logarithms.
As shown in the next section, the ratios of the NLO over the LO
contributions in the new perturbative expansions are smaller
than the ratios evaluated in the conventional approach.

%%%%%%%%%%%%%%%%%%%%%%%%%%%%%%%%%%%%%%%%%%%%%%%%%%%%%%%%%%%%%%%%%%%%%%%%%
\section{Numerical Results and Predictions}
\label{sec5}

In this section, we present numerical evaluations of the leading 
and next-to-leading order Drell-Yan cross section at large
transverse momentum. We show the quantitative differences between the
predictions of the modified factorization formula of Eq.~(\ref{DY-mfac}) 
and the conventional factorization formula.  We demonstrate the sensitivity 
of the cross section to the gluon distribution at low $Q^2$ and high $Q_T$. 

We employ the CTEQ5M set of
parton distributions \cite{CTEQ5}.  We use a two-loop expression for
the strong coupling strength $\alpha_s$, with the value of
$\Lambda_{\rm QCD}$ specified by CTEQ5M, and a one loop expression for
$\alpha_{em}$ with $\alpha_{em}(M_Z)=1/128$.  We choose $Q^2$ as the
renormalization scale for $\alpha_{em}$.  We equate the
renormalization and factorization scales and set 
hard scale $\mu=\mu_f=\kappa\sqrt{Q_T^2+Q^2}$ with 
constant $\kappa=O(1)$.  

The normal $\overline{MS}$ factorization scheme 
removes the ultraviolet $1/\epsilon$ pole of the parton-level
fragmentation functions along with the corresponding splitting functions for
massless partons, but the scheme does not guarantee that the fragmentation 
functions to a massive parton or to a photon with non-vanishing invariant mass
\cite{QZ-VPFF} will be positive.  As a result, the virtual photon fragmentation
functions calculated in the $\overline{MS}$ scheme can be negative in the 
region of 
large $z$~\cite{QZ-VPFF,BL-frag}.  As long as the cross section
is positive, a negative fragmentation function is simply a
particular separation of finite contributions between the
coefficient function and the fragmentation function. It is not a 
problem in principle. Nevertheless, it is more appealing intuitively that 
the fragmentation functions be positive definite.  Since the virtual 
photon fragmentation
functions are purely perturbative, it is possible to preserve 
positivity of the fragmentation functions if we require that the
mass threshold constraints be respected at every stage of the
fragmentation (or bremsstrahlung radiation).  

As shown in Ref.~\cite{QZ-VPFF}, the ultraviolet (UV) divergences of
the virtual photon fragmentation functions come from two 
sources: (1) elementary divergent diagrams associated with the
renormalization of the fields and coupling constants in QCD and QED,
and (2) the loop momenta of the skeleton ladder diagrams.  An
invariant mass cutoff scheme is introduced in Ref.~\cite{QZ-VPFF} 
to render the fragmentation functions positive definite.
In this scheme, all UV divergences associated with the internal
elementary divergent diagrams are removed in the same way as the QCD 
and QED Lagrangian are renormalized (say, in the $\overline{MS}$
scheme).  The UV divergences connected with loop momenta of
the skeleton ladder diagrams and wave function renormalization of
the composite operators are removed by imposition of an invariant mass cut
on the loop momenta.  All running coupling constants in
this scheme are renormalized in the same way as the renormalization of 
the Lagrangian.  Use of the same invariant mass cutoff on the loop momenta of
the skeleton ladder diagrams and the virtual diagrams due to the wave
function renormalization of the composite operators ensures the
infrared cancellation between the real and the virtual diagrams
\cite{CQ-evo}.  In this invariant mass cut-off scheme, a
parton-to-virtual-photon fragmentation function
$D_{c\rightarrow\gamma^*(Q) X}(z,\mu_F^2;Q^2)$ can be viewed as an
inclusive rate for ``decay'' of the parton of flavor $c$ and squared 
invariant mass $\mu_F^2$ into a virtual photon of squared invariant mass
$Q^2$ and momentum fraction $z$ \cite{QZ-VPFF}.    

\begin{figure}
\begin{center}
\epsfig{figure=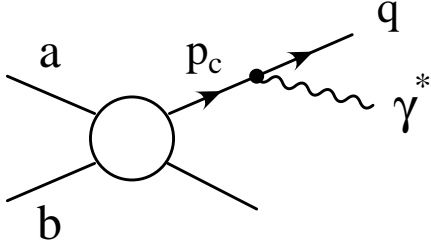,width=2.5in}
\end{center}
\caption{A generic diagram of the lowest order $2\rightarrow 3$
subprocess that contributes to the Drell-Yan cross section with
large final-state logarithmic terms when $Q^2\ll Q_T^2$.} 
\label{fig3}
\end{figure}

\begin{figure}
\begin{center}
\epsfig{figure=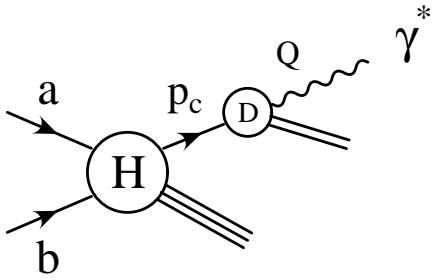,width=2.5in}
\end{center}
\caption{Sketch of the fragmentation contribution to low mass
Drell-Yan lepton-pair production at high $Q_T$.}
\label{fig4}
\end{figure}

\begin{figure}
\begin{center}
\epsfig{figure=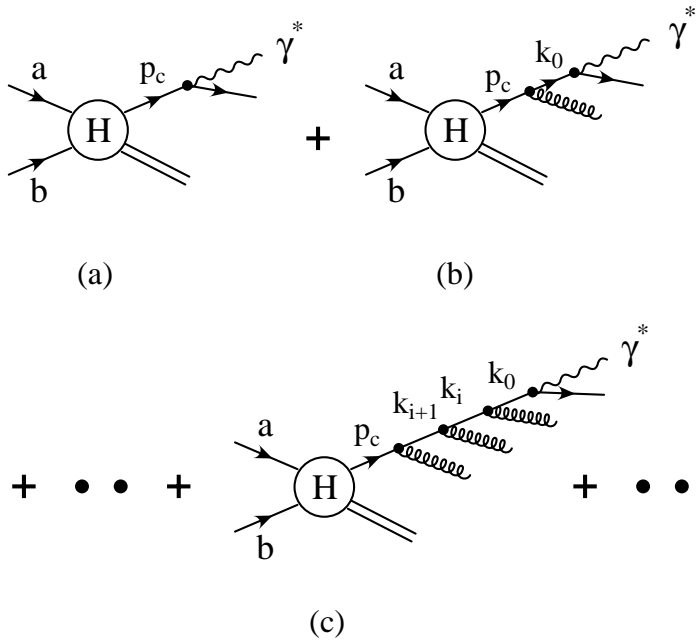,width=3.8in}
\end{center}
\caption{Scattering amplitudes that provide large logarithmic 
contributions to the Drell-Yan cross section via quark
fragmentation.} 
\label{fig5}
\end{figure}

\begin{figure}
\begin{center}
\epsfig{figure=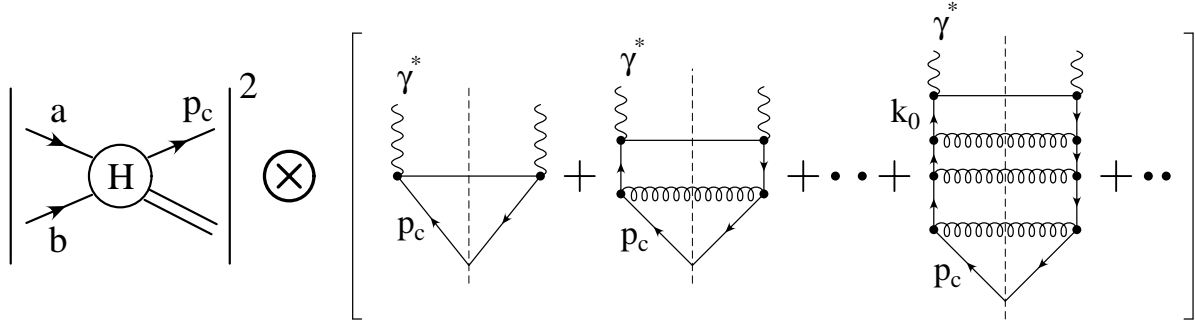,width=6.2in}
\end{center}
\caption{Factored non-singlet quark-to-virtual-photon fragmentation
contributions to the Drell-Yan cross section.}
\label{fig6}
\end{figure}

In our numerical calculations, we use the virtual photon
fragmentation functions in the invariant mass cut-off scheme of 
Ref.~\cite{QZ-VPFF}. The invariant mass for
quark $c$ in Fig.~\ref{fig3} to decay into a virtual photon of invariant 
mass $Q$ and a massless quark is 
\begin{equation}
p_c^2 = \frac{1}{z(1-z)}\, \vec{Q}_T^2 + \frac{Q^2}{z}\, .
\label{pc2}
\end{equation}
The three-vector $\vec{Q}_T$ is perpendicular to the direction of the 
parent quark's momentum.  The fragmentation
scale is chosen to be the invariant mass of the fragmenting parton.  For
example, we choose $\mu_F^2 = p_c^2$ at lowest order. 
Equation~(\ref{pc2}) shows that the mass threshold requires that the
fragmentation scale be $\mu_F^2\ge Q^2/z$ \cite{QZ-VPFF,BL-frag}.
If we keep 
the perturbative contributions to a high enough order, the 
cross section should not be sensitive to the choice of
fragmentation scale.  However, since the fragmentation scale in this
new scheme is very different from the traditional scale in 
the $\overline{MS}$ scheme, we test two choices for 
the fragmentation scale,
\begin{eqnarray}
\mu_F &=& \kappa\, \sqrt{Q_T^2+Q^2}\, ,
\label{frag-s-1} \\
\mu_F &=& \kappa\, \sqrt{\frac{Q_T^2+Q^2}{z}}\, .
\label{frag-s-2}
\end{eqnarray}
The first choice is the same as that for the renormalization and
factorization scales.  

The second choice in Eq.~(\ref{frag-s-2}) is motivated by the fact
that the squared invariant mass of the fragmenting parton is of the
order $(Q_T^2+Q^2)/z$, obtained as follows.  For the
generic $2\rightarrow 3$ diagram in Fig.~\ref{fig3}, with the final 
parton that recoils against $p_c$ assumed massless, we compute 
\begin{eqnarray}
p_c^2 &=& 2(p_a+p_b)\cdot p_c - (p_a+p_b)^2
\nonumber \\
&\approx & - \frac{t+u}{z}\, .
\label{pc2-in}
\end{eqnarray}
To derive the second line, we use the approximations 
$Q^2\ll Q_T^2$ and $s < |t+u|/z$.  If we let $\mu_F^2\sim p_c^2$,
Eq.~(\ref{pc2-in}) leads to the fragmentation logarithm,
\begin{equation}
\ln\left(\frac{\mu_F^2}{Q^2/z}\right) 
\approx \ln\left(\frac{-(t+u)/z}{Q^2/z}\right) 
=  \ln\left(\frac{-(t+u)}{Q^2}\right) \, ,
\label{log-frag}
\end{equation}
consistent with the logarithm in Eq.~(\ref{log-M}). 
Since $|t+u|\sim O(Q_T^2+Q^2)$, we find $\mu_F^2 \sim (Q_T^2+Q^2)/z$.

The perturbatively calculated partonic parts in conventional
perturbation theory,
$\hat{\sigma}_{ab\rightarrow\gamma^*(Q) X}^{(C-LO)}$ 
and  $\hat{\sigma}_{ab\rightarrow\gamma^*(Q) X}^{(C-NLO)}$ in
Eq.~(\ref{DY-m-LO-NLO-2}), are available and
calculated in the $\overline{\rm MS}$ scheme \cite{BGK-DY,AR-DY}. The
fragmentation contribution in Eq.~(\ref{DY-F-LO}) depends on the
leading order short-distance partonic parts, 
$\hat{\sigma}_{ab\rightarrow c X}^{(F-LO)}$, and on the virtual photon
fragmentation functions.  The partonic parts are 
\begin{equation}
\frac{\hat{\sigma}_{ab\rightarrow c X}^{(F-LO)}}{dp_{c_T}^2\,dy}
= \frac{\pi}{2 x_1 x_2 S}
\left|\frac{1}{g^2} \overline{M}_{ab\rightarrow c X} \right|^2\,
\left(8\pi^2\right)\,
\delta\left((x_1P_A+x_2P_B-p_c)^2\right) 
\left(\frac{\alpha_s(\mu)}{2\pi}\right)^2\, .
\label{Habc2}
\end{equation}
The functions $\overline{M}_{ab\rightarrow c X}$ are the lowest order
matrix elements for partons $a$ and $b$ to produce a parton
of flavor $c$, averaged over the colors and spins of the partons in the 
initial state.  They are available in Ref.~\cite{Owens-RMP}.

To compare with data, we introduce the invariant differential cross section 
\begin{eqnarray}
E\frac{d^3\sigma_{AB\rightarrow \ell^+\ell^- X}}{d^3Q}
&=& \frac{1}{\pi} \int dQ^2 \left[
\frac{d\sigma_{AB\rightarrow \ell^+\ell^-(Q) X}}{dQ^2\,dQ_T^2\,dy}
\right]
\nonumber\\
&=& \frac{\alpha_{em}}{3\pi^2}\,
    \int \frac{dQ^2}{Q^2} \left[
    \frac{d\sigma_{AB\rightarrow \gamma^*(Q) X}}{dQ_T^2\,dy}
    \right] .
\label{DY-diff}
\end{eqnarray}
The integration in $dQ^2$ is over a bin centered on the invariant mass $Q$. 
The differential cross sections  
$d\sigma_{AB\rightarrow\ell^+\ell^- X}/dQ^2\,dQ_T^2\,dy$ and 
$d\sigma_{AB\rightarrow\gamma^* X}/dQ_T^2\,dy$ are given in  
Eq.~(\ref{DY-Vph}) and Eq.~(\ref{DY-m-LO-NLO}), respectively.  If the
bin size $\Delta Q$ is much
smaller than $Q$, the cross section can be approximated as 
\begin{equation}
E\frac{d^3\sigma_{AB\rightarrow \ell^+\ell^- X}}{d^3Q}
\approx 
    \frac{2\alpha_{em}}{3\pi^2}\,
    \frac{\Delta Q}{Q}\, \left[
    \frac{d\sigma_{AB\rightarrow \gamma^*(Q) X}}{dQ_T^2\,dy} 
    \right] .
\label{DY-approx}
\end{equation}
The differential cross section through next-to-leading order 
for production of a virtual photon, 
$d\sigma_{AB\rightarrow \gamma^*(Q) X}/dQ_T^2\,dy$, is given in
Eq.~(\ref{DY-m-LO-NLO}). 

Since we do not present the NLO partonic hard parts for the
fragmentation contributions in this paper, we use 
Eq.~(\ref{DY-m-LO-NLO}) without the $d\hat{\sigma}_{ab\rightarrow
c X}^{(F-NLO)}/dp_{c_T}^2 dy$ term for the numerical
calculations of the Drell-Yan cross sections.

\begin{figure}
\begin{center}
\begin{minipage}[t]{3in}
\epsfig{figure=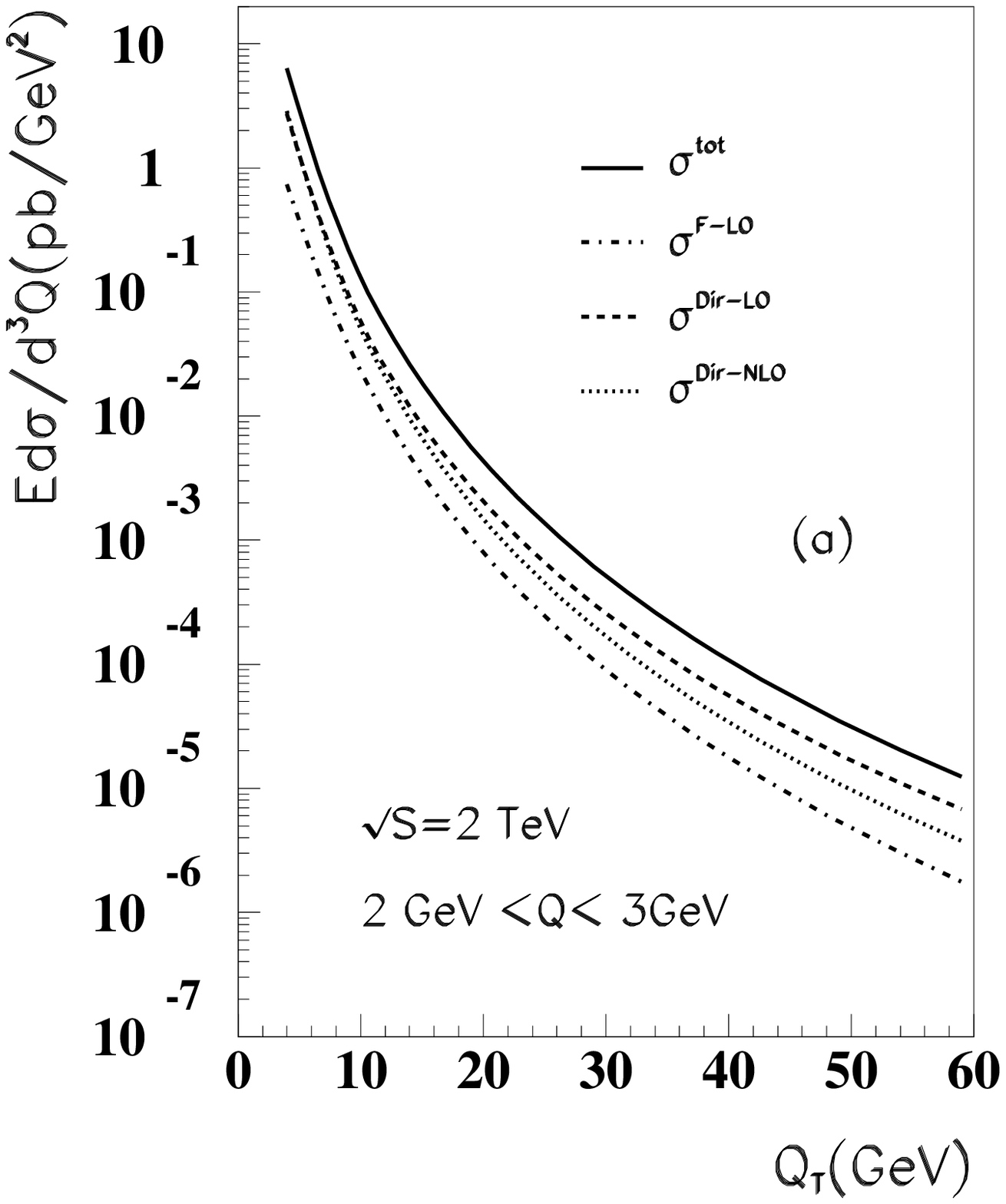,width=3.0in}
\end{minipage}
\hfill
\begin{minipage}[t]{3in}
\epsfig{figure=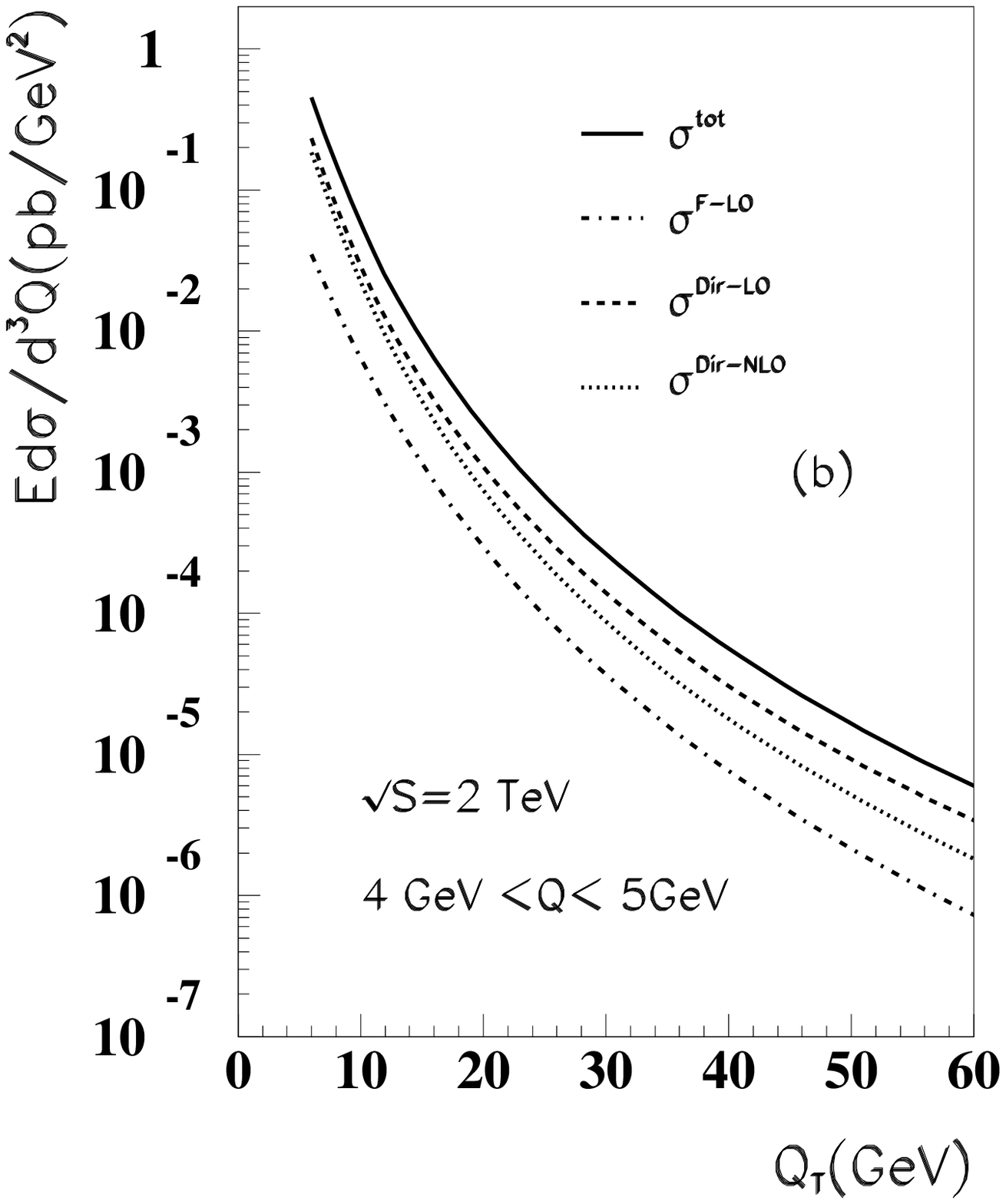,width=3.0in}
\end{minipage}
\end{center}
\caption{Drell-Yan cross section as a function of $Q_T$ at the Tevatron
energy $\sqrt{S}=2.0$~TeV and rapidity $y=0$ for the mass intervals  
(a) $2\le Q\le 3$~GeV and (b) $4\le Q\le 5$~GeV.  
Solid, dashed, dotted, and dot-dashed lines stand for the total, LO
direct, NLO direct, and resummed fragmentation contributions,
respectively. }
\label{fig7}
\end{figure}

\begin{figure}
\begin{center}
\begin{minipage}[t]{3in}
\epsfig{figure=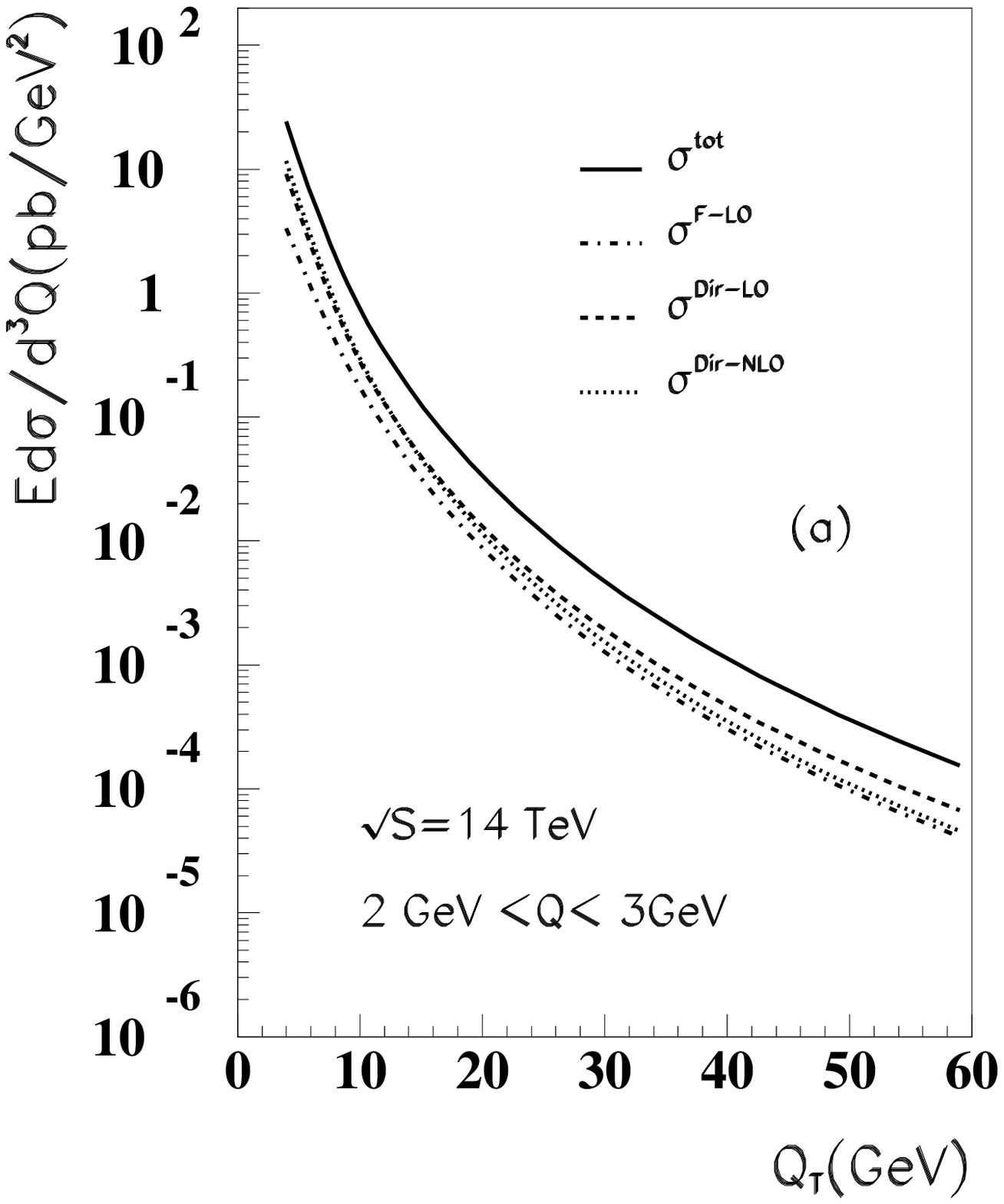,width=2.6in}
\end{minipage}
\hfill
\begin{minipage}[t]{3in}
\epsfig{figure=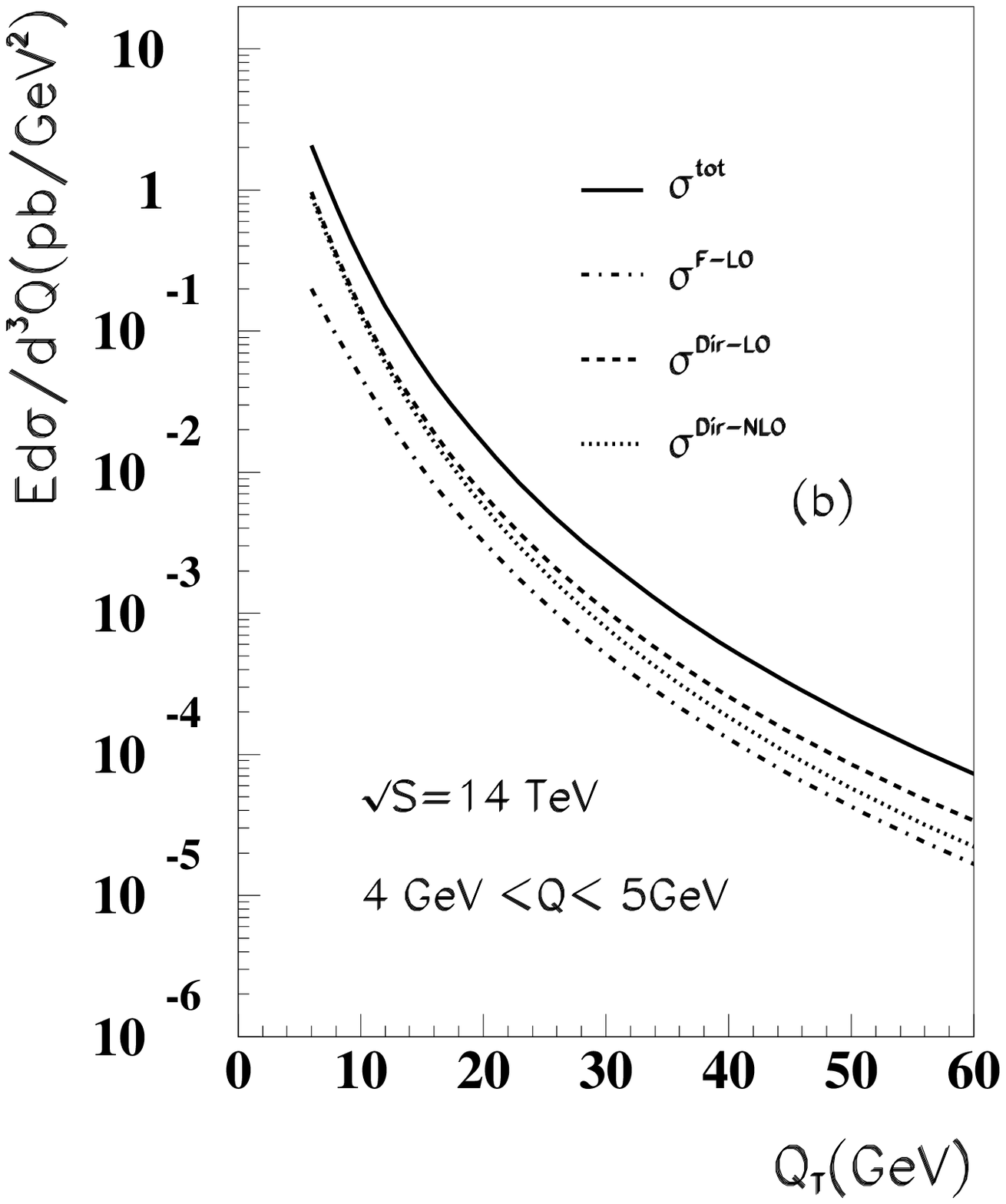,width=2.6in}
\end{minipage}
\end{center}
\caption{Drell-Yan cross section as a function of $Q_T$ at the LHC
energy $\sqrt{S}=14$~TeV and rapidity $y=0$ for the mass intervals  
(a) $2\le Q\le 3$~GeV and (b) $4\le Q\le 5$~GeV.  
Solid, dashed, dotted, and dot-dashed lines stand for the total, LO
direct, NLO direct, and resummed fragmentation contributions,
respectively. }
\label{fig8}
\end{figure}

\begin{figure}
\begin{center}
\begin{minipage}[t]{3in}
\epsfig{figure=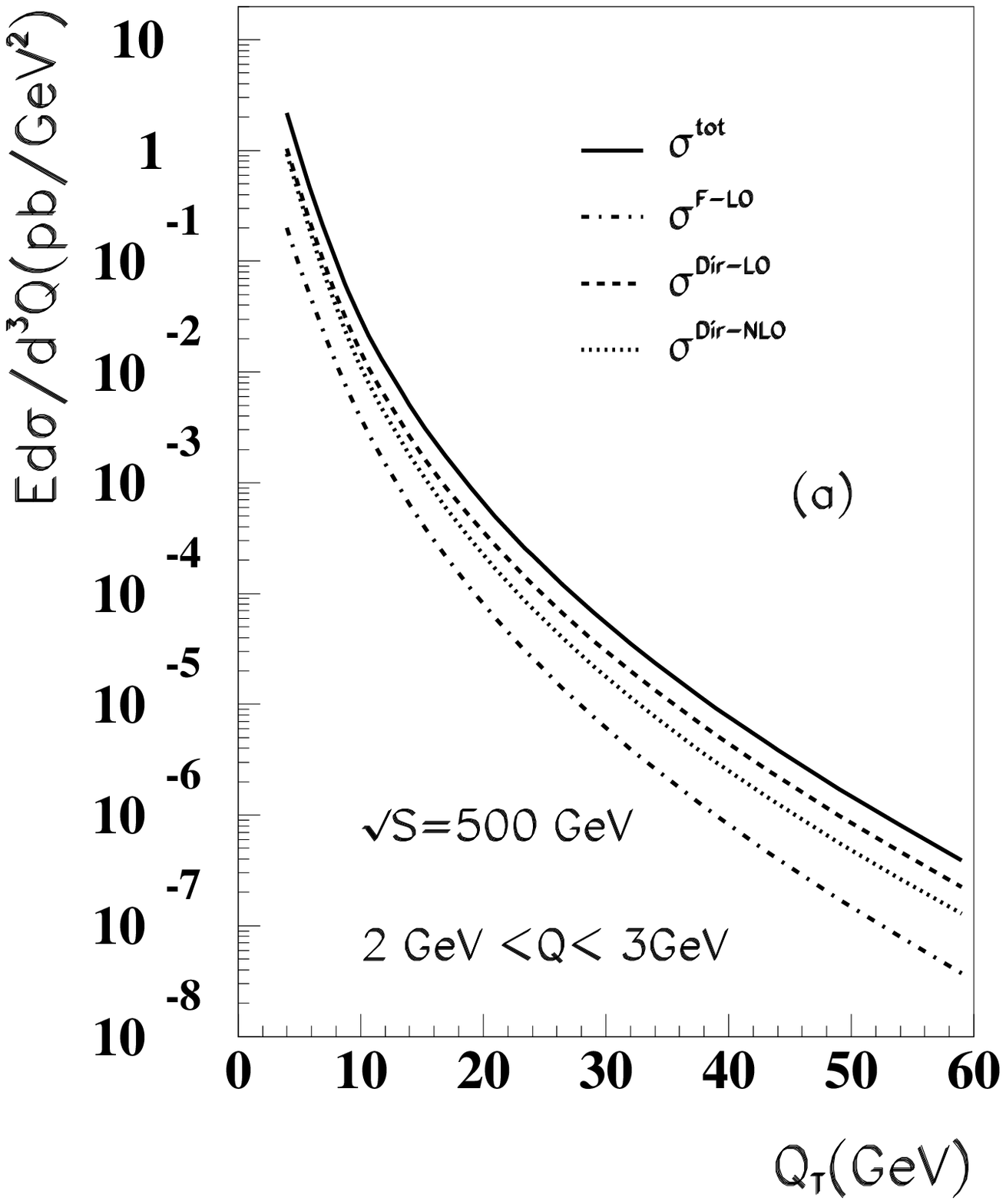,width=2.6in}
\end{minipage}
\hfill
\begin{minipage}[t]{3in}
\epsfig{figure=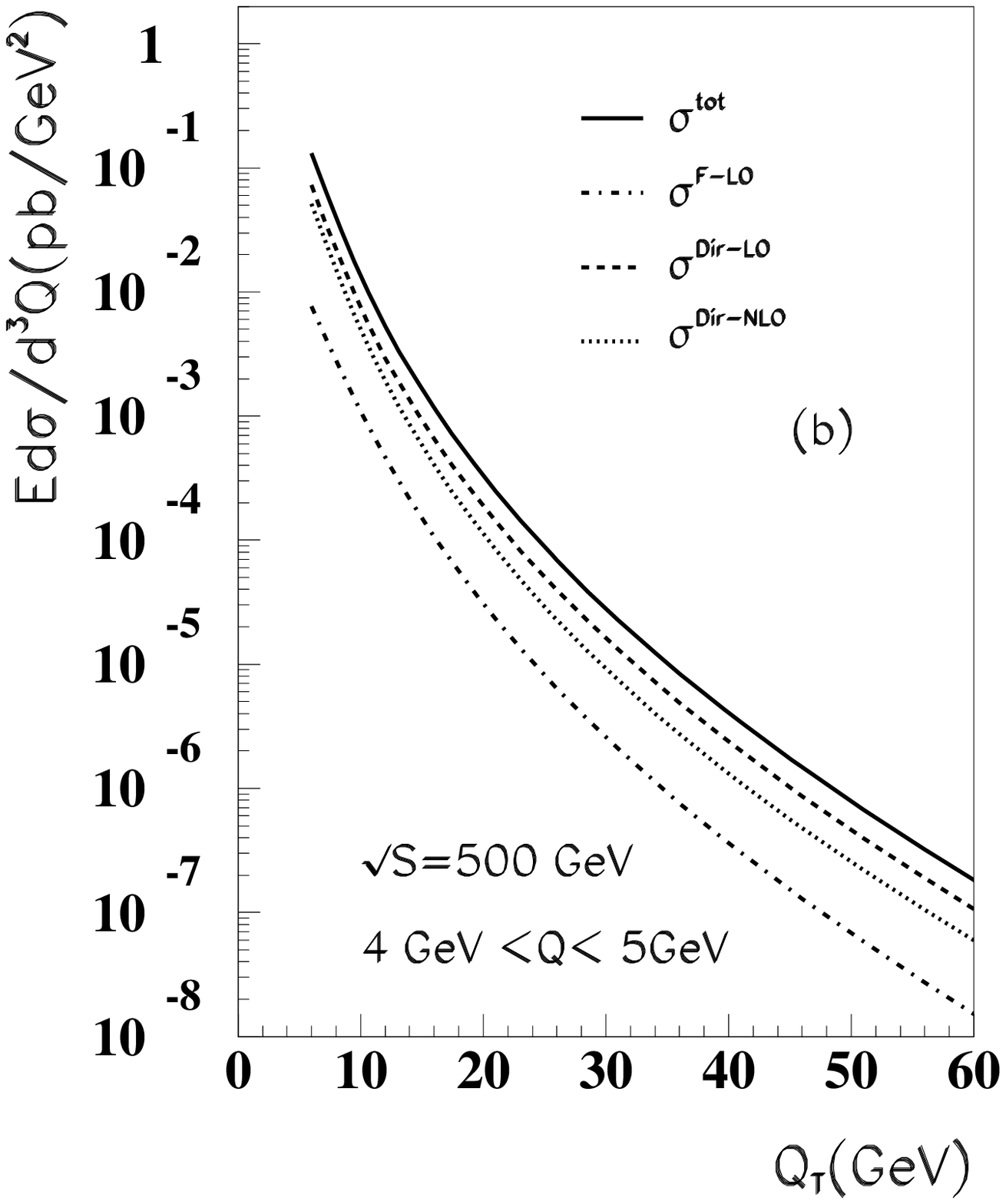,width=2.6in}
\end{minipage}
\end{center}
\caption{Drell-Yan cross section as a function of $Q_T$ at the RHIC
proton-proton energy $\sqrt{S}=500$~GeV and rapidity $y=0$ for the mass 
intervals (a) $2\le Q\le 3$~GeV and (b) $4\le Q\le 5$~GeV.  
Solid, dashed, dotted, and dot-dashed lines stand for the total, LO
direct, NLO direct, and resummed fragmentation contributions,
respectively. }
\label{fig9}
\end{figure}

In Fig.~\ref{fig7}, \ref{fig8}, and \ref{fig9}, we plot the 
differential cross section $Ed^3\sigma/d^3Q$ as a 
function of $Q_T$ for $p\bar{p}\rightarrow \ell^+\ell^- +X$ 
at the Tevatron collider energy $\sqrt{S}=2.0$~TeV, for $pp\rightarrow
\ell^+\ell^- +X$ at the LHC energy $\sqrt{S}=14$~TeV, and for    
$pp\rightarrow \ell^+\ell^- +X$ at the RHIC proton-proton energy
$\sqrt{S}=500$~GeV.  In all three figures, the rapidity $y=0$.  The 
renormalization, factorization, and fragmentation scales are set equal, 
with constant $\kappa=1$.
The solid, dashed, dotted, and dot-dashed lines represent
the total, the LO direct, the NLO direct, and the resummed
fragmentation contributions, respectively.  We show results for 
virtual photon invariant mass $2\le Q\le 3$~GeV and $4\le Q\le 5$~GeV.  
(Note: the solid curve represents the sum of the dashed, dotted, and 
dot-dashed lines.)  To enhance the cross sections, the smaller
value of the invariant mass $Q$ is favored as long as the lepton pairs 
can be identified experimentally.

At the integrated luminosities of run I of the Fermilab collider, prompt 
real photons have been observed with values of transverse momentum extending 
to 100 GeV and beyond~\cite{tevdat}.  Scaling to the massive lepton pair case, 
we judge that it should be possible to examine cross sections in the same 
data sample out to $Q_T$ of 30~GeV or more for virtual photon 
invariant mass $2\le Q\le 3$~GeV.  Values of $Q_T$ to 50~GeV or so may 
be reached with $2 {\rm fb}^{-1}$ at run II.  

\begin{figure}
\begin{center}
\begin{minipage}[t]{3in}
\epsfig{figure=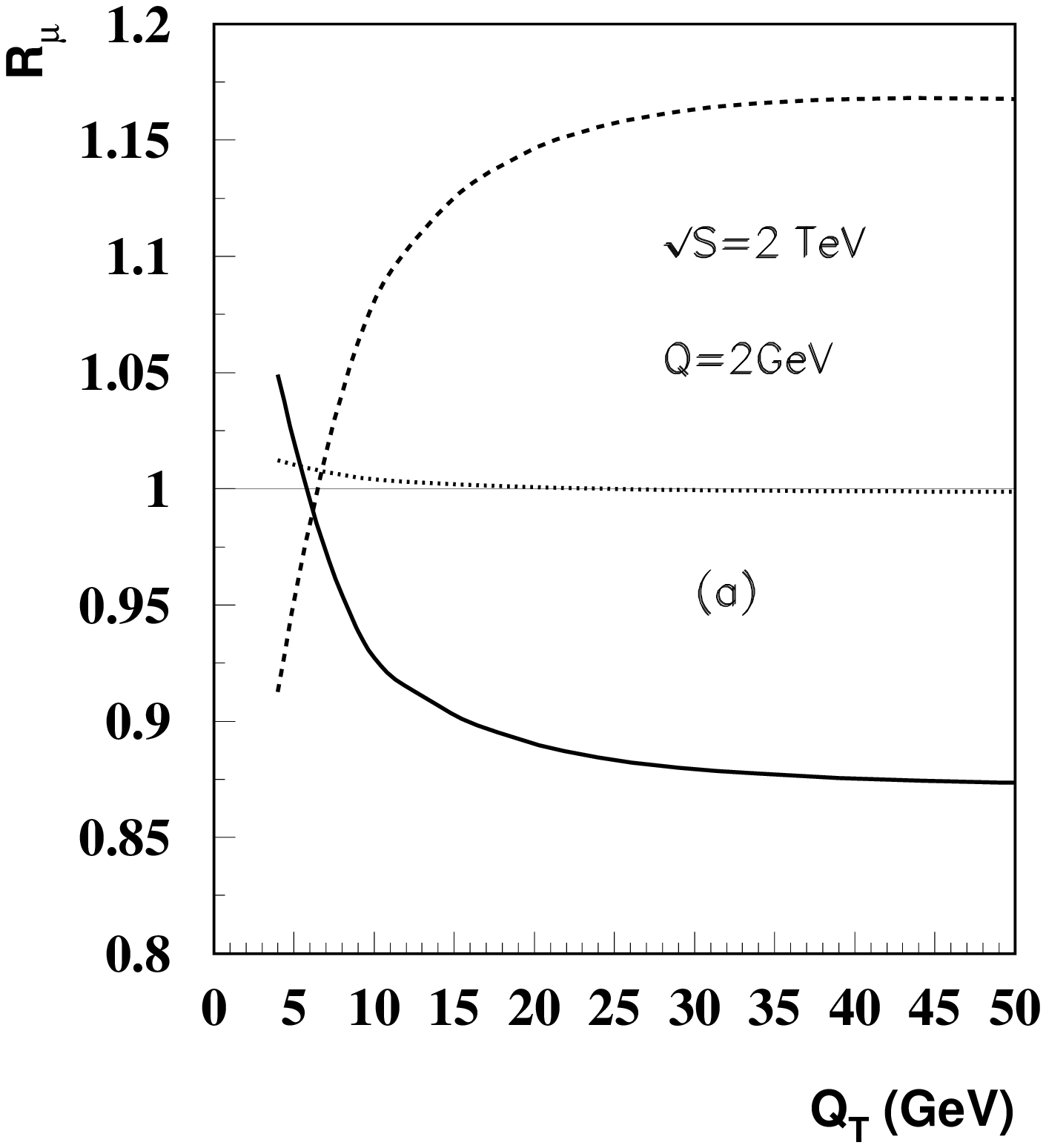,width=3.0in}
\end{minipage}
\hfill
\begin{minipage}[t]{3in}
\epsfig{figure=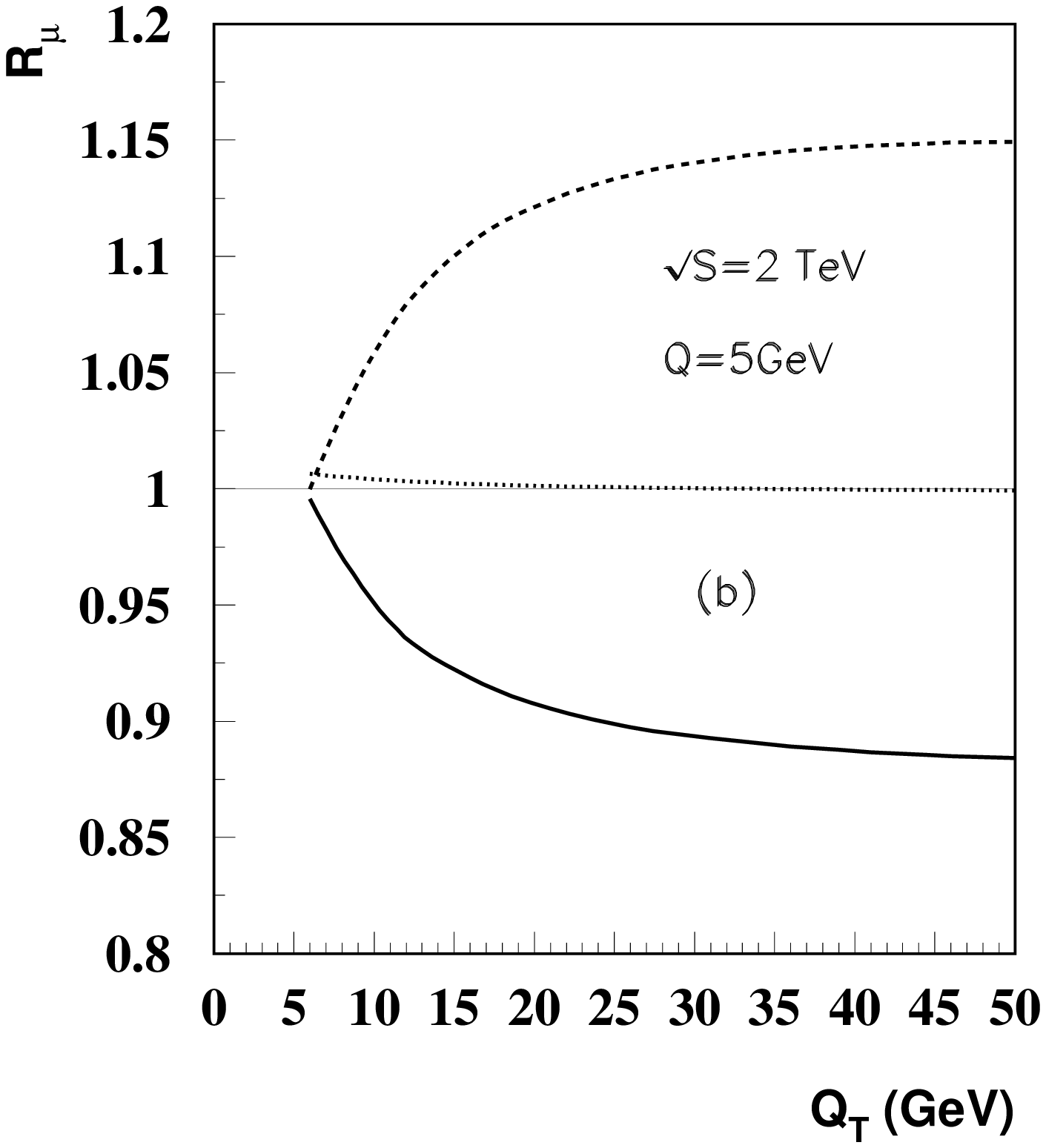,width=3.0in}
\end{minipage}
\end{center}
\caption{Ratio $R_\mu$ in Eq.~(\protect\ref{R-mu}) as a function of
$Q_T$ at $y=0$ and the Tevatron energy $\protect\sqrt{S}=2$~TeV for 
virtual photon invariant mass (a) $2$~GeV and (b) $5$~GeV.  The curves 
are explained in the text.}
\label{fig10}
\end{figure}

To demonstrate the renormalization, factorization and
fragmentation scale-dependence of the cross sections, we introduce the 
ratio
\begin{equation}
R_{\mu} \equiv
\left.
 E\frac{d{\sigma}_{AB\rightarrow \ell^+\ell^- X}}
     {d^3Q} (\kappa)
\right/
 E\frac{d{\sigma}_{AB\rightarrow \ell^+\ell^- X}}
     {d^3Q} (\kappa=1) \, .  
\label{R-mu}
\end{equation}
The denominator in Eq.~(\ref{R-mu}) is obtained with all 
three scales equal: $\mu=\mu_f=\mu_F=\sqrt{Q_T^2+Q^2}$.   
In Fig.~\ref{fig10}, we plot $R_\mu$ as a function of $Q_T$
at the Tevatron energy and $y=0$, for $2$~GeV and $5$~GeV, with
different scale choices.      
We first fix all three scales to be the same:
$\mu=\mu_f=\mu_F=\kappa\sqrt{Q_T^2+Q^2}$, with $\kappa=2$ (solid) and
$\kappa=1/2$ (dashed).  The dotted lines correspond to the choice
$\mu=\mu_f=\sqrt{Q_T^2+Q^2}$, and $\mu_F=\sqrt{(Q_T^2+Q^2)/z}$, as
defined in Eq.~(\ref{frag-s-2}).  As shown in Fig.~\ref{fig10}, the 
scale dependence is not great.  It yields an uncertainty of about 15 \%
for all reasonable values of $Q_T$ at the Tevatron energy.
The effect of the different choice for the
fragmentation scale in Eq.~(\ref{frag-s-2}) is also small.  
The same features are preserved at the other collider
energies. 

To show the quantitative difference between the conventional
fixed-order perturbative calculations and the calculations with 
all-orders resummation of the large logarithmic terms, we introduce the 
ratios  
\begin{equation}
R^{(LO)} \equiv \left.\left[
 E\frac{d{\sigma}_{AB\rightarrow\ell^+\ell^- X}^{(Dir-LO)}}{d^3Q} 
+E\frac{d{\sigma}_{AB\rightarrow\ell^+\ell^- X}^{(F-LO)}}{d^3Q} 
\right] \right/
\left[
 E\frac{d{\sigma}_{AB\rightarrow\ell^+\ell^- X}^{(C-LO)}}{d^3Q} 
\right]
\label{R-LO}
\end{equation}
for the LO contributions, and 

\begin{equation}
R\equiv \left.\left[
 E\frac{d{\sigma}_{AB\rightarrow\ell^+\ell^- X}^{(Dir-LO)}}{d^3Q} 
+E\frac{d{\sigma}_{AB\rightarrow\ell^+\ell^- X}^{(Dir-NLO)}}{d^3Q} 
+E\frac{d{\sigma}_{AB\rightarrow\ell^+\ell^- X}^{(F-LO)}}{d^3Q} 
\right] \right/
\left[
 E\frac{d{\sigma}_{AB\rightarrow\ell^+\ell^- X}^{(C-LO)}}{d^3Q} 
+E\frac{d{\sigma}_{AB\rightarrow\ell^+\ell^- X}^{(C-NLO)}}{d^3Q} 
\right]\, .
\label{R}
\end{equation}

\begin{figure}
\begin{center}
\begin{minipage}[t]{3in}
\epsfig{figure=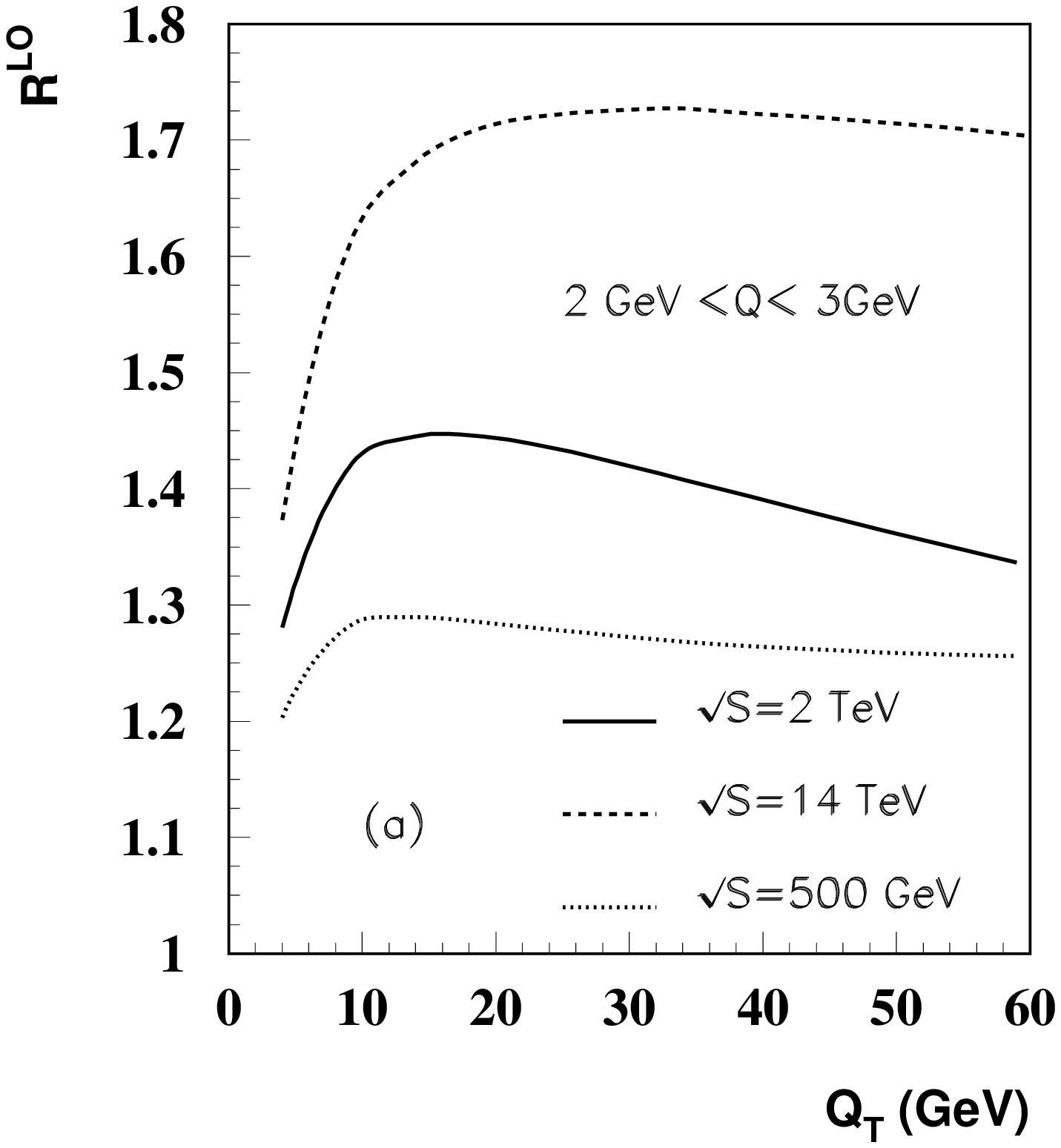,width=3.0in}
\end{minipage}
\hfill
\begin{minipage}[t]{3in}
\epsfig{figure=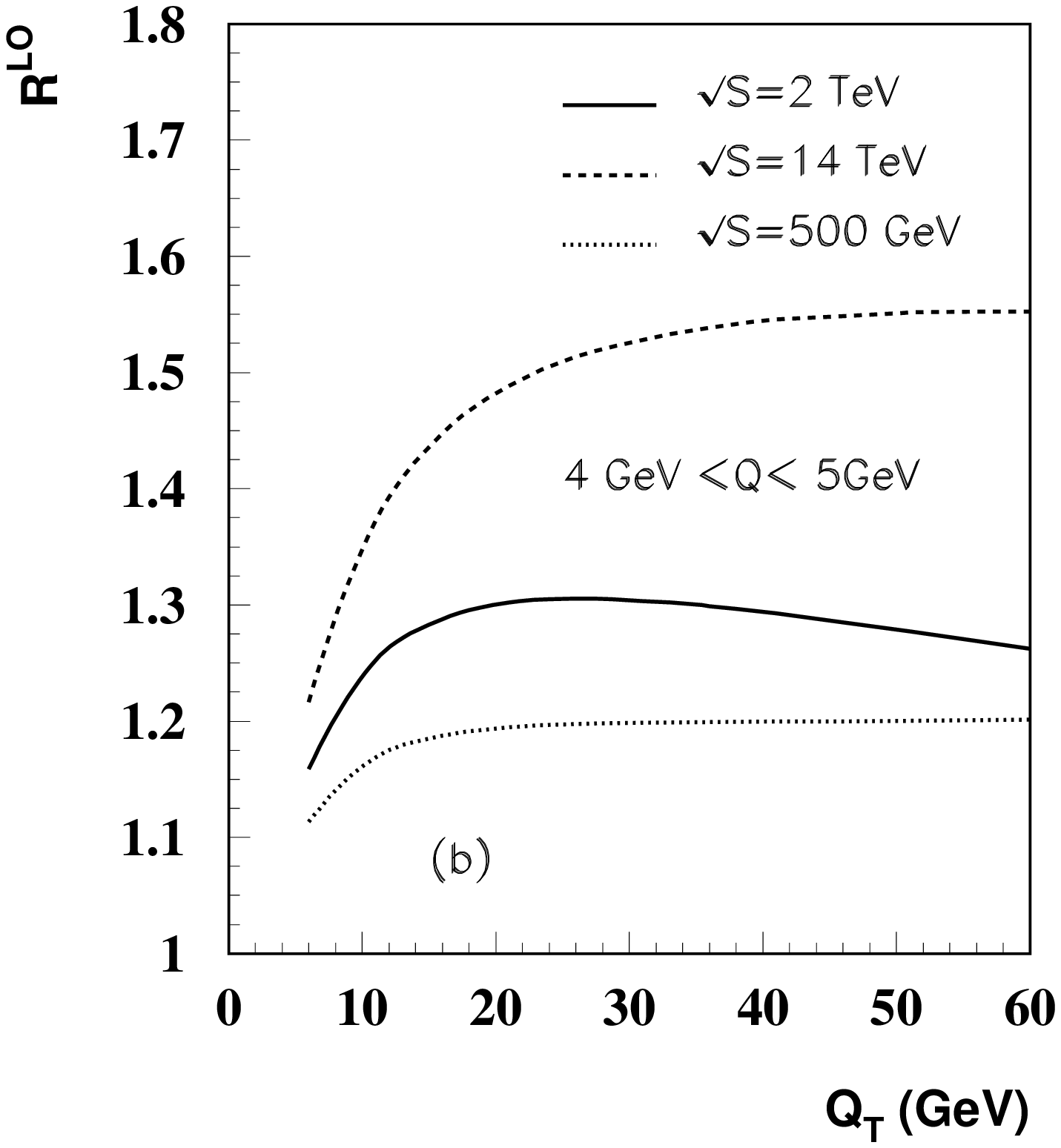,width=3.0in}
\end{minipage}
\end{center}
\caption{Ratio $R^{(LO)}$ in Eq.~(\protect\ref{R-LO}) as a function of
$Q_T$ at $y=0$ and three different collision energies for the mass intervals 
(a) $2\le Q\le 3$~GeV and (b) $4\le Q\le 5$~GeV.  The curves are explained
in the text.}
\label{fig11}
\end{figure}

The ratio $R^{(LO)}$ is the ratio of the LO contributions in 
the two different factorization formalisms.  Since the LO direct
contribution is the same as the LO term in the conventional
calculation, any deviation of $R^{(LO)}$ from unity measures the
size of the logarithmic contributions.  In Fig.~\ref{fig11}, we
plot the ratio $R^{(LO)}$ as a function of $Q_T$ at different
energies for $2\le Q\le 3$~GeV and $4\le Q\le
5$~GeV.  The solid, dashed, and dotted lines represent the Tevatron,
LHC, RHIC energies.  Owing to the threshold behavior of the fragmentation 
function, the fragmentation contribution vanishes for $Q_T \sim Q$, and 
$R^{(LO)} \rightarrow 1$.  The logarithmic
contributions are very important at the LHC energy and less
important at RHIC energies.  The reason for this energy dependence 
is the phase space penalty associated with the large invariant mass of 
the virtual photon.   The large logarithm in the
fragmentation function is proportional to $\ln(z\, \mu_F^2/Q^2)$ 
for each power of $\alpha_s$.  Since the strong coupling
strength $\alpha_s(\mu)$ is proportional to $
1/\ln(\mu^2/\Lambda_{\rm QCD}^2)$, the product 
\begin{equation}
\alpha_s(\mu)\, \ln(z\, \mu_F^2/Q^2) 
\propto \frac{\ln(z\, \mu_F^2/Q^2)}
             {\ln(\mu^2/\Lambda_{\rm QCD}^2)}\, 
\label{log-power}
\end{equation}
becomes of order unity only when $\mu_F\sim \mu \gg Q^2$.
Otherwise, the combination in Eq.~(\ref{log-power}) is not large
because of the factors $z$ and $Q^2$ from the mass threshold.

The ratio $R$ in Eq.~(\ref{R}) measures the difference between the 
conventional calculation up to NLO accuracy and our resummed
calculation, obtained from our modified
factorization formalism {\it without} the NLO fragmentation
contributions.  Without the NLO fragmentation contributions
$Ed\sigma^{(F-NLO)}/d^3Q$ in the numerator, the ratio $R$ does not
represent the entire ratio of the NLO contributions in the two different
factorization approaches.  Nevertheless, this ratio does indicate some of 
the effects of QCD resummation.  

The difference between the numerator and the denominator of the ratio
$R$ in Eq.~(\ref{R}) is determined entirely by the second term in
Eq.~(\ref{DY-m-LO-NLO-2}) without the NLO fragmentation contributions, 
\begin{eqnarray}
&\ &
\sum_{a,b,c} \int dx_1 \phi_{a/A}(x_1,\mu) 
     \int dx_2 \phi_{b/B}(x_2,\mu)
     \int \frac{dz}{z^2}\, 
     \frac{d\hat{\sigma}_{ab\rightarrow c X}^{(F-LO)}}{dp_{c_T}^2 dy}
     (x_1,x_2,p_c=\frac{\hat{Q}}{z};\mu_F,\mu) 
\nonumber \\
&\ &{\hskip 0.2in}
\times \left[D_{c\rightarrow\gamma^*(Q) X}(z,\mu_F^2;Q^2)
          -D_{c\rightarrow\gamma^*(Q) X}^{(0)}(z,\mu_F^2;Q^2) \right] .
\label{DY-dF-LO}
\end{eqnarray}
This difference is proportional to the difference between the LO QED
parton-to-virtual-photon fragmentation functions 
$D_{c\rightarrow\gamma^*(Q) X}^{(0)}(z,\mu_F^2;Q^2)$ and the corresponding
QCD evolved parton-to-virtual-photon fragmentation functions
$D_{c\rightarrow\gamma^*(Q) X}(z,\mu_F^2;Q^2)$.  As shown in
Ref.~\cite{QZ-VPFF}, one of the major differences between these 
fragmentation functions is the behavior at large $z$:
\begin{eqnarray}
   D_{c\rightarrow\gamma^*(Q) X}^{(0)}(z\rightarrow 1,\mu_F^2;Q^2) 
   &\neq & 0
\nonumber \\
   D_{c\rightarrow\gamma^*(Q) X}(z\rightarrow 1,\mu_F^2;Q^2) 
   &=& 0 \, .
\label{D-z-to-1}
\end{eqnarray}
QCD evolution reduces the fragmentation function at large $z$ while
it increases the fragmentation function at small $z$.  If the cross
section is dominated by the small (large) $z$ region, the numerator of
the ratio $R$ in Eq.~(\ref{R}) is larger (smaller) than the
denominator calculated in fixed-order perturbation theory.  

The $z$-integration of the Drell-Yan cross section runs from
$z_{\rm min}$ to $1$. We introduce a cutoff $z_c$ to limit the
integration to the range $z_{\rm min}$ to $z_c$, and we define the ratio  
\begin{equation}
R_z \equiv 
\left.
 E\frac{d{\sigma}_{AB\rightarrow \ell^+\ell^- X}^{(F-LO)}}
     {d^3Q} (z_c)
\right/
 E\frac{d{\sigma}_{AB\rightarrow \ell^+\ell^- X}^{(F-LO)}}
     {d^3Q} (z_c=1) \, .
\label{R-z}
\end{equation}
The cutoff $z_c$ can be between $z_{\rm min}$ and $1$;  
$R_z(z_c=z_{\rm min})=0$, and $R_z(z_c=1)=1$.  The shape of the
ratio $R_z$ establishes which region of $z$ dominates
the $z$-integration.

\begin{figure}
\begin{center}
\epsfig{figure=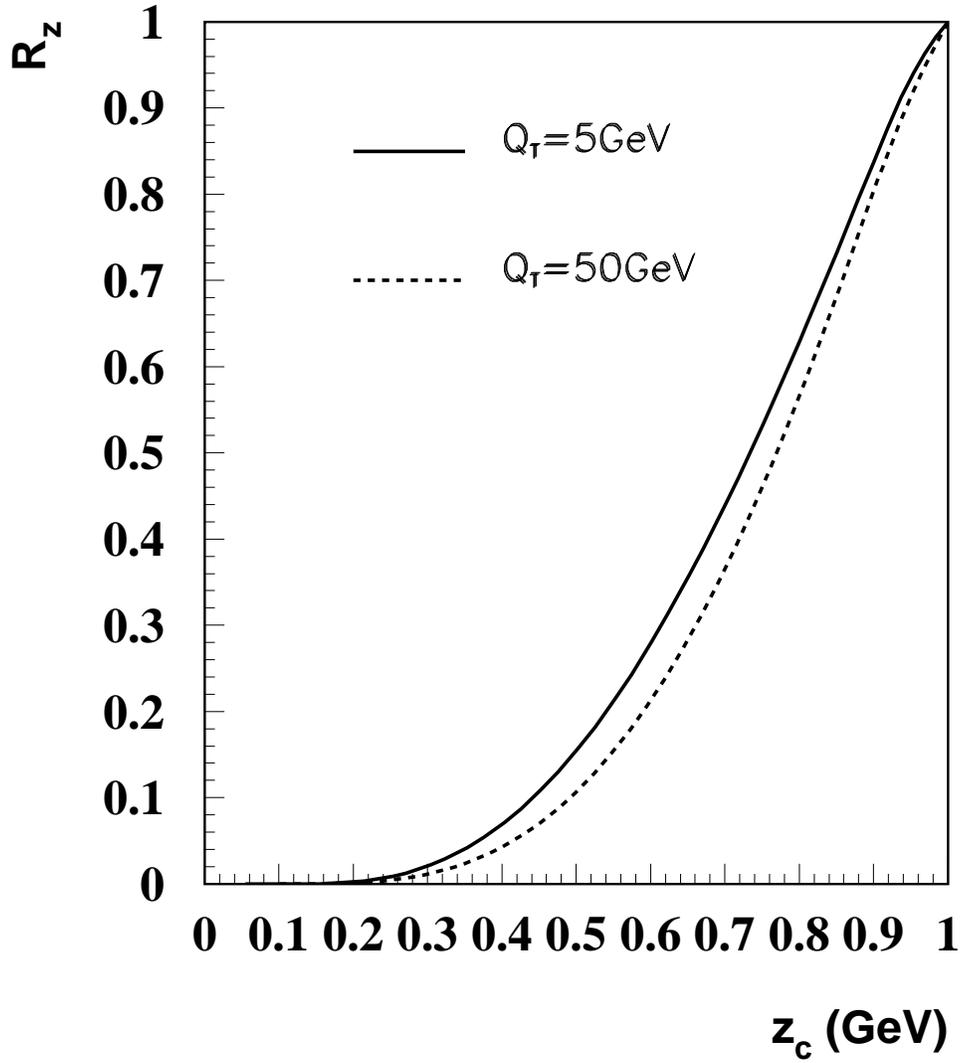,width=5.0in}
\end{center}
\caption{Ratio $R_z$ in Eq.~(\protect\ref{R-z}) as a function of
$z_c$ at the Tevatron energy $\protect\sqrt{S}=2$~TeV, $y=0$, and
$Q=2$~GeV.  The solid and dashed lines correspond to transverse 
momenta $Q_T=5$ and 50~GeV.} 
\label{fig12}
\end{figure}

In Fig.~\ref{fig12}, we plot $R_z$ as a
function of $z_c$ at the Tevatron energy with $Q=2$~GeV.  The 
solid and dashed lines correspond to transverse momenta 
$Q_T=5$ and 50~GeV.  If the integrand for the
$z$-integration were independent of $z$, the
ratio $R_z$ would be proportional to $(z_c-z_{\rm min})/(1-z_{\rm
min})$, which corresponds to a straight line for $R_z$.  
The shape of $R_z$ in Fig.~\ref{fig12} shows that  
the $z$-integration is dominated by the large $z$ region.

Since the $z$-dependence of the partonic cross section in
Eq.~(\ref{DY-F-LO}) tends to cancel the $1/z^2$ factor in the
$z$-integration, the $z$-integration for the Drell-Yan cross section
is determined mainly by the convolution of parton distributions and
fragmentation functions and by the shapes of these functions.  

For a given collision energy $\sqrt{S}$, the transverse momentum $Q_T
= z\, p_{c_T} \propto z\, x\, \sqrt{S}$ with $x\sim x_1\sim x_2$ for 
$y \simeq 0$.  For a 
fixed value of $Q_T$, a large (small) $z$ corresponds to the small
(large) $x$ region.  Because the parton distributions and
fragmentation functions are both steeply falling functions of the
momentum fractions ($x$ or $z$), and the Drell-Yan cross section is
proportional to two powers of parton distributions and one power of
fragmentation function, the convolution of parton distributions and
fragmentation functions favors the combination of a small $x$ and large
$z$.  The fragmentation contributions are dominated by the
small $x$ and large $z$ region.  The $z$-integration over the difference 
between the QCD
evolved fragmentation function and the leading order QED fragmentation
function in Eq.~(\ref{DY-dF-LO}) is dominated by the large $z$
region.  The net contribution from this term is expected to
be negative, and the ratio $R$ in Eq.~(\ref{R}) should be less than
one for $Q_T \gg Q$.  

\begin{figure}
\begin{center}
\begin{minipage}[t]{3in}
\epsfig{figure=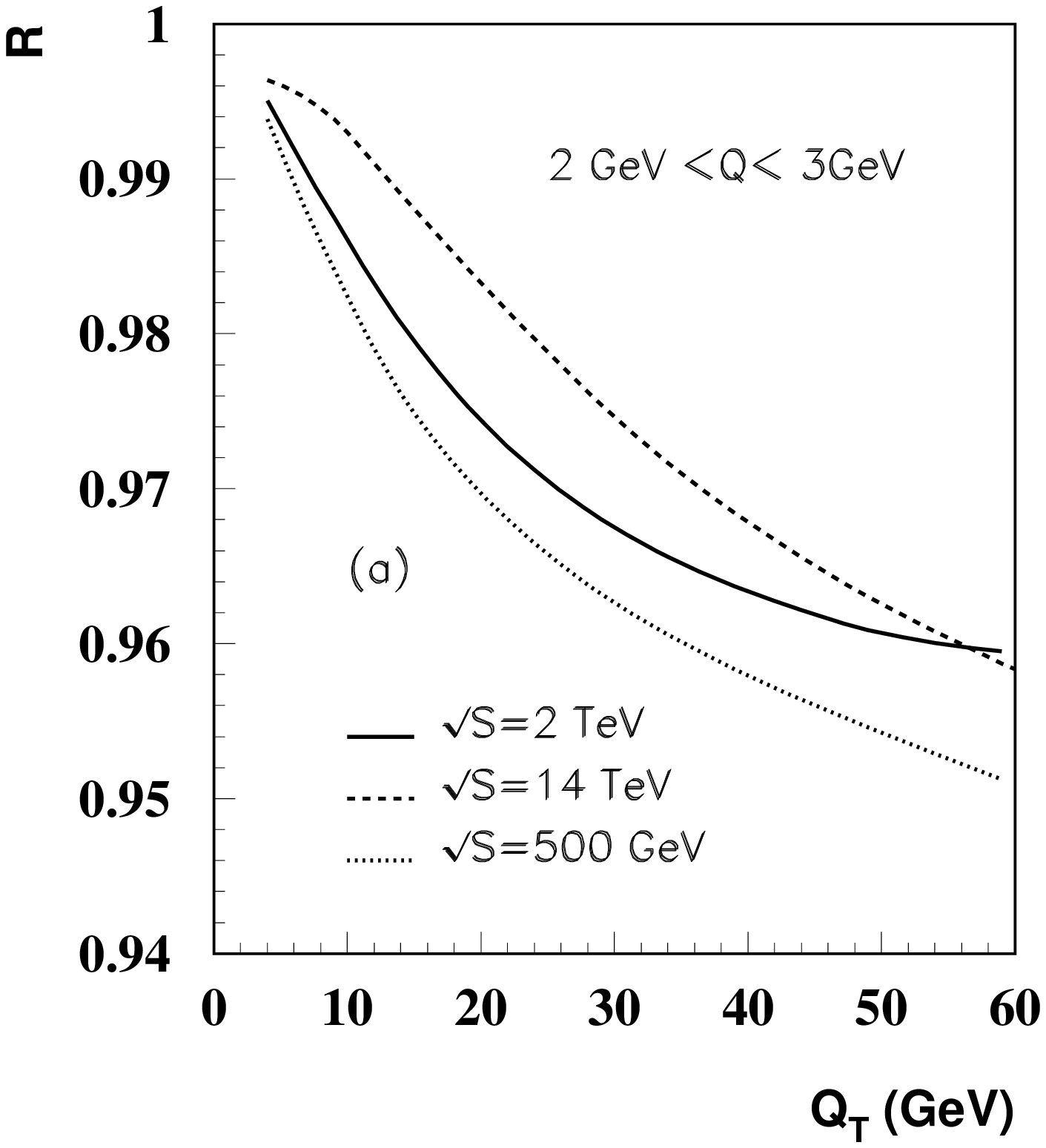,width=3.0in}
\end{minipage}
\hfill
\begin{minipage}[t]{3in}
\epsfig{figure=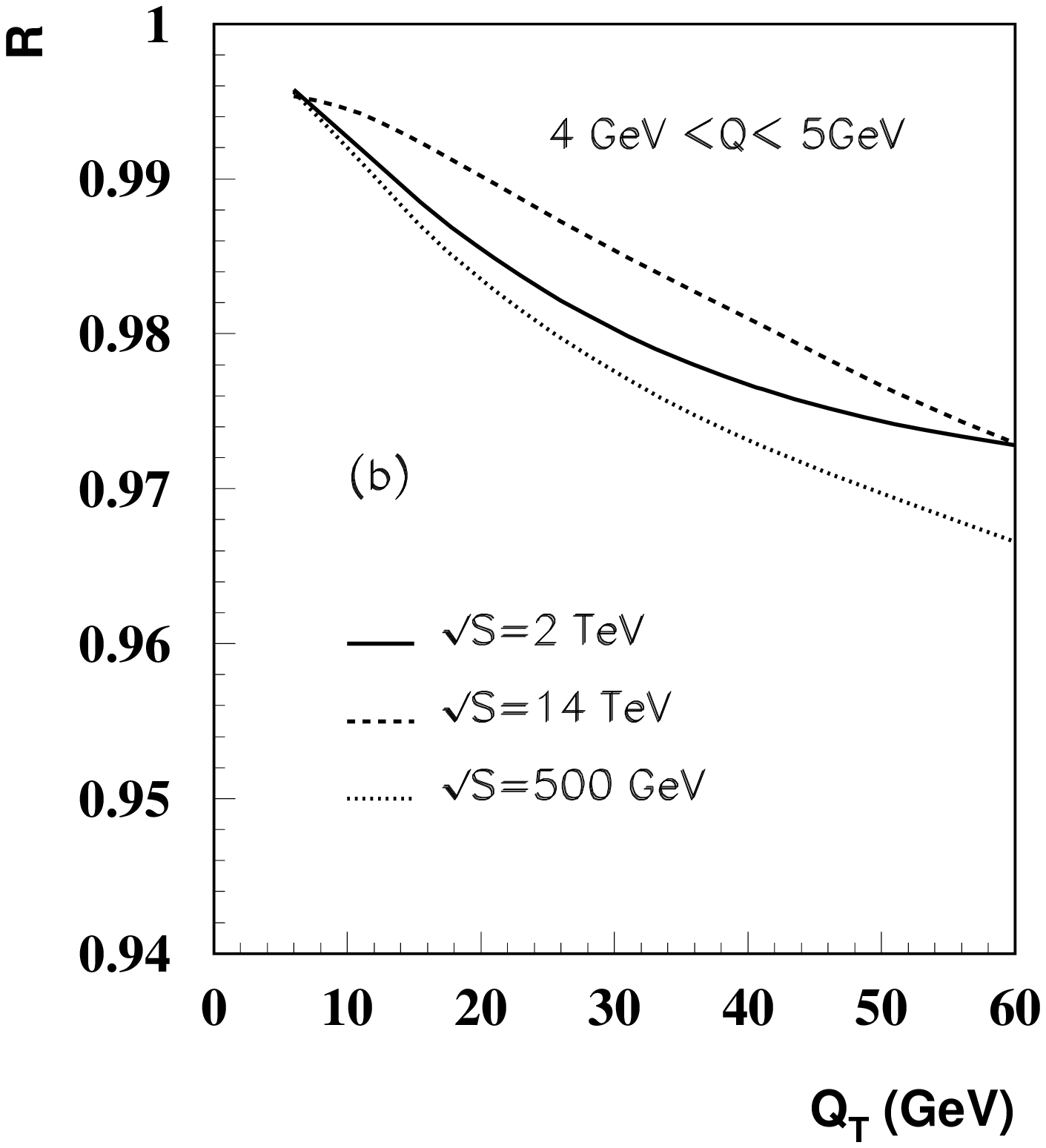,width=3.0in}
\end{minipage}
\end{center}
\caption{Ratio $R$ in Eq.~(\protect\ref{R}) as a function of $Q_T$ at
$y=0$ and mass intervals (a) $2\le Q\le 3$~GeV and (b) $4\le Q\le 5$~GeV.  
Solid,
dashed, and dotted lines are for the Tevatron energy $\sqrt{S}=2.0$~TeV,
the LHC energy $\sqrt{S}=14$~TeV, and the RHIC proton-proton energy  
$\sqrt{S}=500$~GeV, respectively. }
\label{fig13}
\end{figure}

We plot the ratio $R$ in Fig.~\ref{fig13} as a function of
$Q_T$ at $y=0$ for $2\le Q\le 3$~GeV and $4\le Q\le 5$~GeV at
different collision energies.  The solid, dashed, and dotted lines in
Fig.~\ref{fig13} correspond to the Tevatron energy $\sqrt{S}=2.0$~TeV,
the LHC energy $\sqrt{S}=14$~TeV, and the RHIC proton-proton energy
$\sqrt{S}=500$~GeV, respectively. The ratio $R$ is less than one when 
$Q_T > Q$, as expected.  As $Q_T$ grows,  
$z_{\rm min}$ becomes larger, and the fragmentation contributions are
dominated by even larger values of $z$.  Consequently, $R$
is a decreasing function of the transverse momentum $Q_T$.  The predicted 
distribution in $Q_T$ is steepened relative to the conventional 
expectation, with the steepening being greater for smaller $\sqrt S$. The 
shape of the ratio $R$ represents the effect of QCD evolution on the 
virtual photon fragmentation functions.  As $Q_T \rightarrow Q$, 
$R \rightarrow 1$, in keeping with the expectation that our modified 
factorization formula should yield the same cross section as conventional 
fixed-order perturbation theory when $Q_T \sim Q$.  The apparent slight 
difference of $R$ from $1$ near $Q_T = Q$ may be attributed to the choice 
of fragmentation scale $\mu_F$.  If we were to set $\mu_F^2 = Q_T^2/z$, our 
result would be the same as the conventional one when $Q_T=Q$.  

The main advantages of our modified factorization formula in
Eq.~(\ref{DY-mfac}) are that the large logarithms are removed from the
coefficient functions of conventional factorization and all 
perturbatively calculated coefficient functions are evaluated at a
single hard scale.  The reliability of the perturbative calculations 
is enhanced.  In order to demonstrate this benefit, we introduce the ratios
\begin{eqnarray}
R_{c} &\equiv & \left.
E\frac{d{\sigma}_{AB\rightarrow\ell^+\ell^- X}^{(C-NLO)}}
      {d^3Q} \right/
E\frac{d{\sigma}_{AB\rightarrow\ell^+\ell^- X}^{(C-LO)}}
      {d^3Q} \, ,
\label{R-con} \\
R_{Dir} &\equiv & \left.
E\frac{d{\sigma}_{AB\rightarrow\ell^+\ell^- X}^{(Dir-NLO)}}
      {d^3Q} \right/
E\frac{d{\sigma}_{AB\rightarrow\ell^+\ell^- X}^{(Dir-LO)}}
      {d^3Q} \, ,
\label{R-dir} \\
R_{F} &\equiv & \left.
E\frac{d{\sigma}_{AB\rightarrow\ell^+\ell^- X}^{(F-NLO)}}
      {d^3Q} \right/
E\frac{d{\sigma}_{AB\rightarrow\ell^+\ell^- X}^{(F-LO)}}
      {d^3Q} \, .
\label{R-frag}
\end{eqnarray}
The subscripts $c$, $Dir$, and $F$ represent the
conventional perturbative calculation, our direct contribution,
and our fragmentation contribution, respectively.  Because the large 
logarithms are removed in the perturbatively calculated coefficient 
functions in the modified factorization formula, we expect the ratios 
$R_{Dir}$ and $R_{F}$ to be smaller than the ratio $R_c$.

\begin{figure}
\begin{center}
\begin{minipage}[t]{3in}
\epsfig{figure=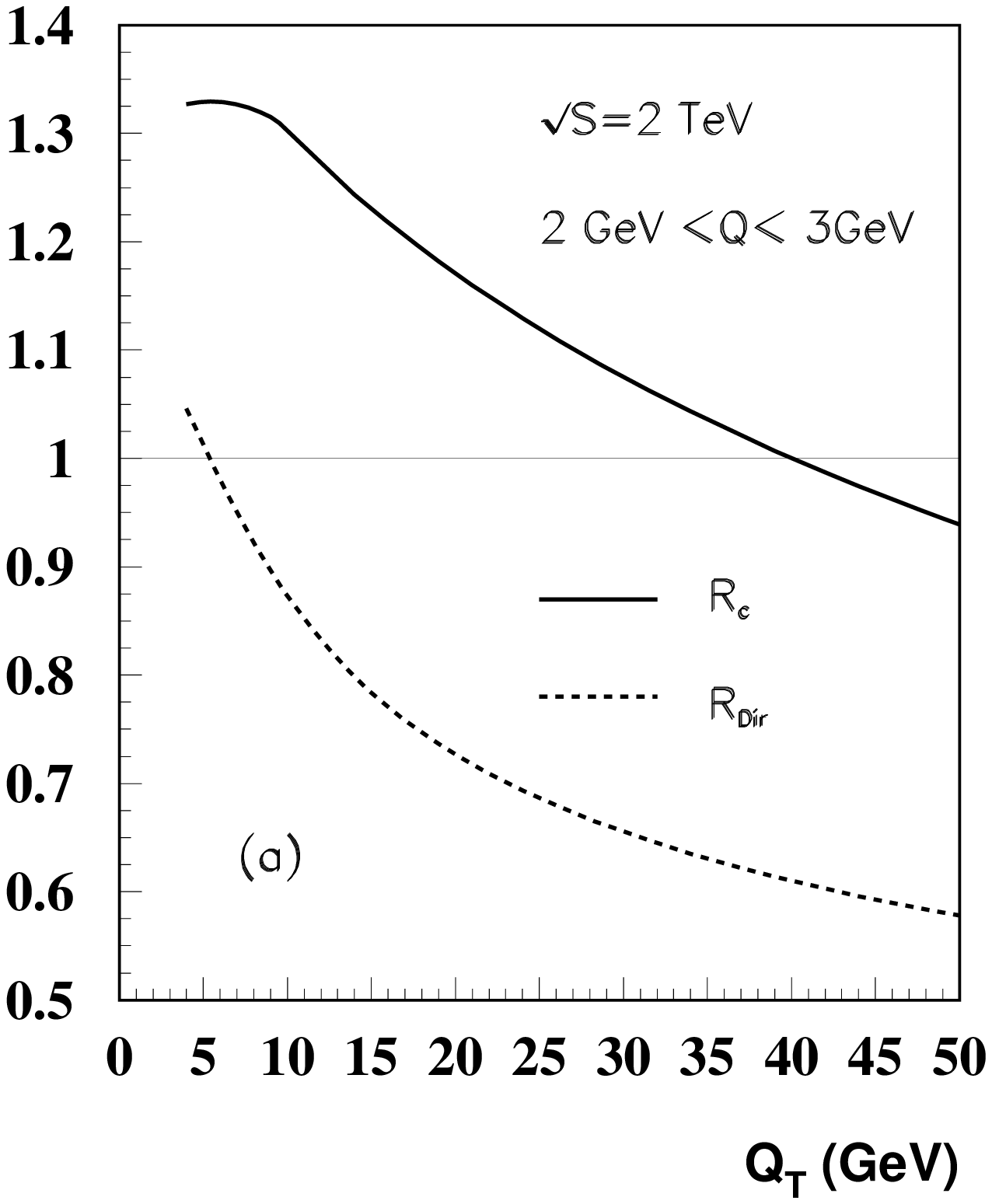,width=3.0in}
\end{minipage}
\hfill
\begin{minipage}[t]{3in}
\epsfig{figure=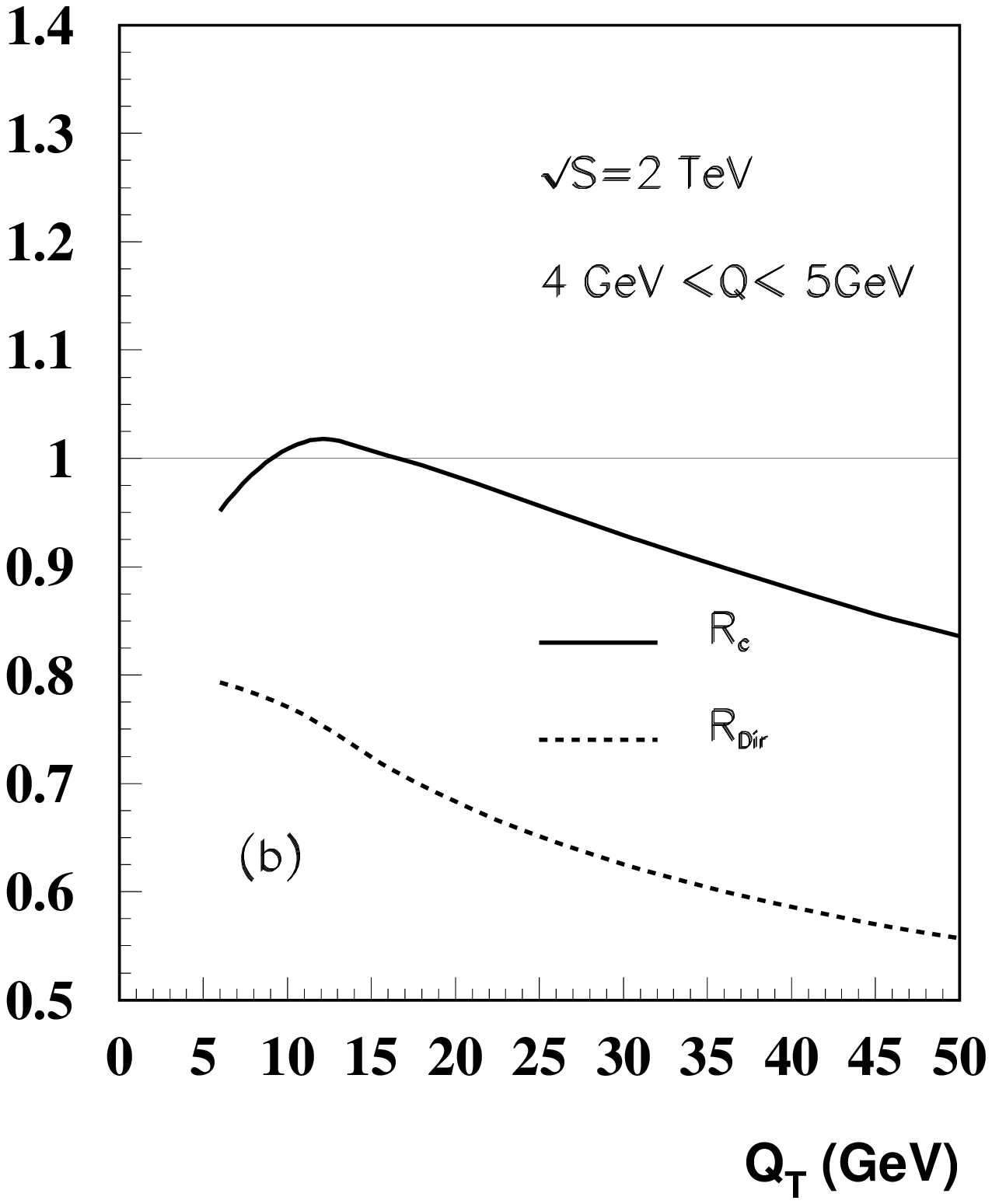,width=3.0in}
\end{minipage}
\end{center}
\caption{The ratios in Eqs.~(\protect\ref{R-con}) and
(\protect\ref{R-dir}) as a function of $Q_T$ at
$y=0$ and for the Tevatron energy $\sqrt{S}=2.0$~TeV and mass 
intervals (a) $2\le Q\le
3$~GeV and (b) $4\le Q \le 5$~GeV.  Solid and dashed lines are for $R_c$
and $R_{Dir}$, respectively.  }
\label{fig14}
\end{figure}

\begin{figure}
\begin{center}
\begin{minipage}[t]{3in}
\epsfig{figure=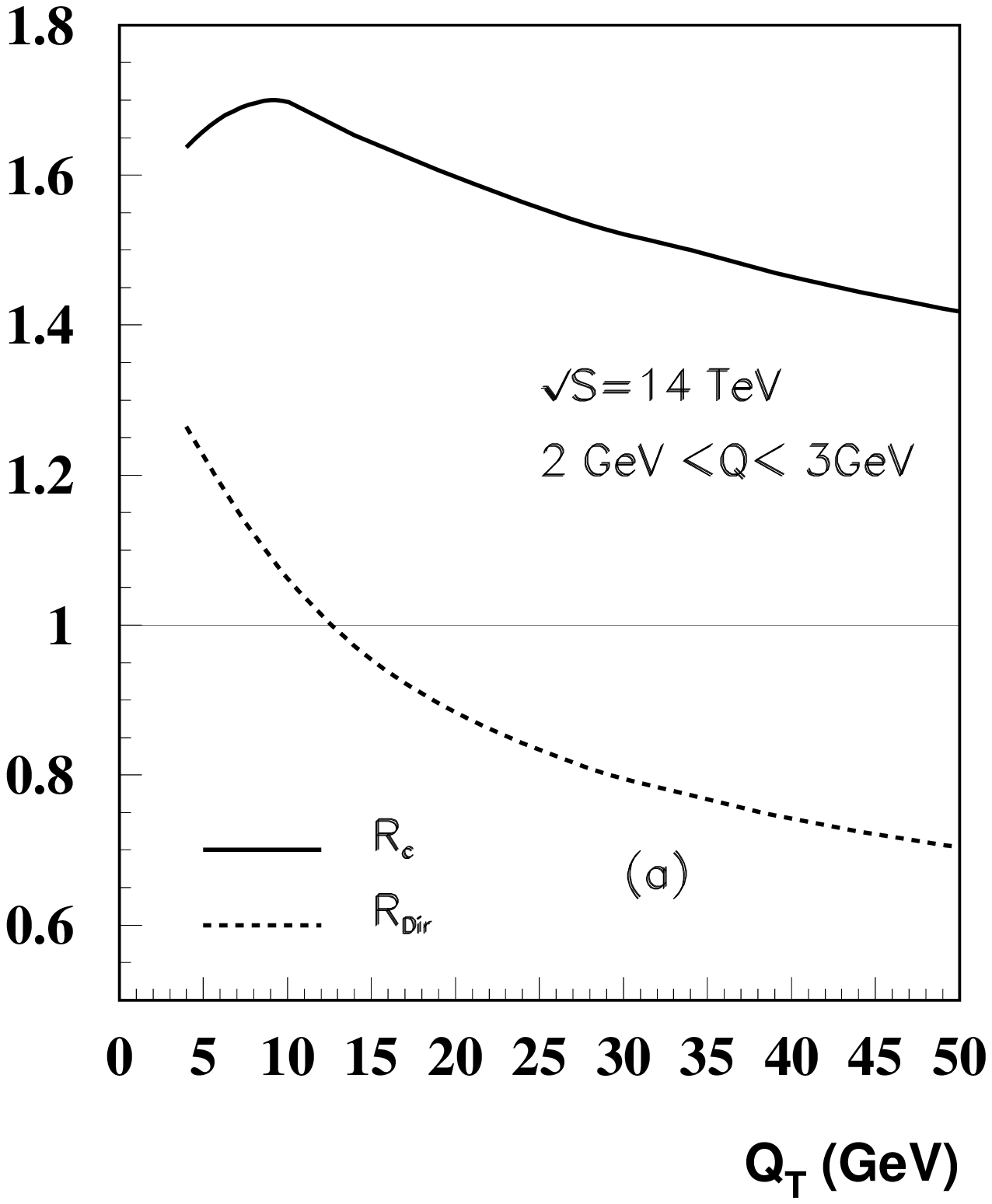,width=3.0in}
\end{minipage}
\hfill
\begin{minipage}[t]{3in}
\epsfig{figure=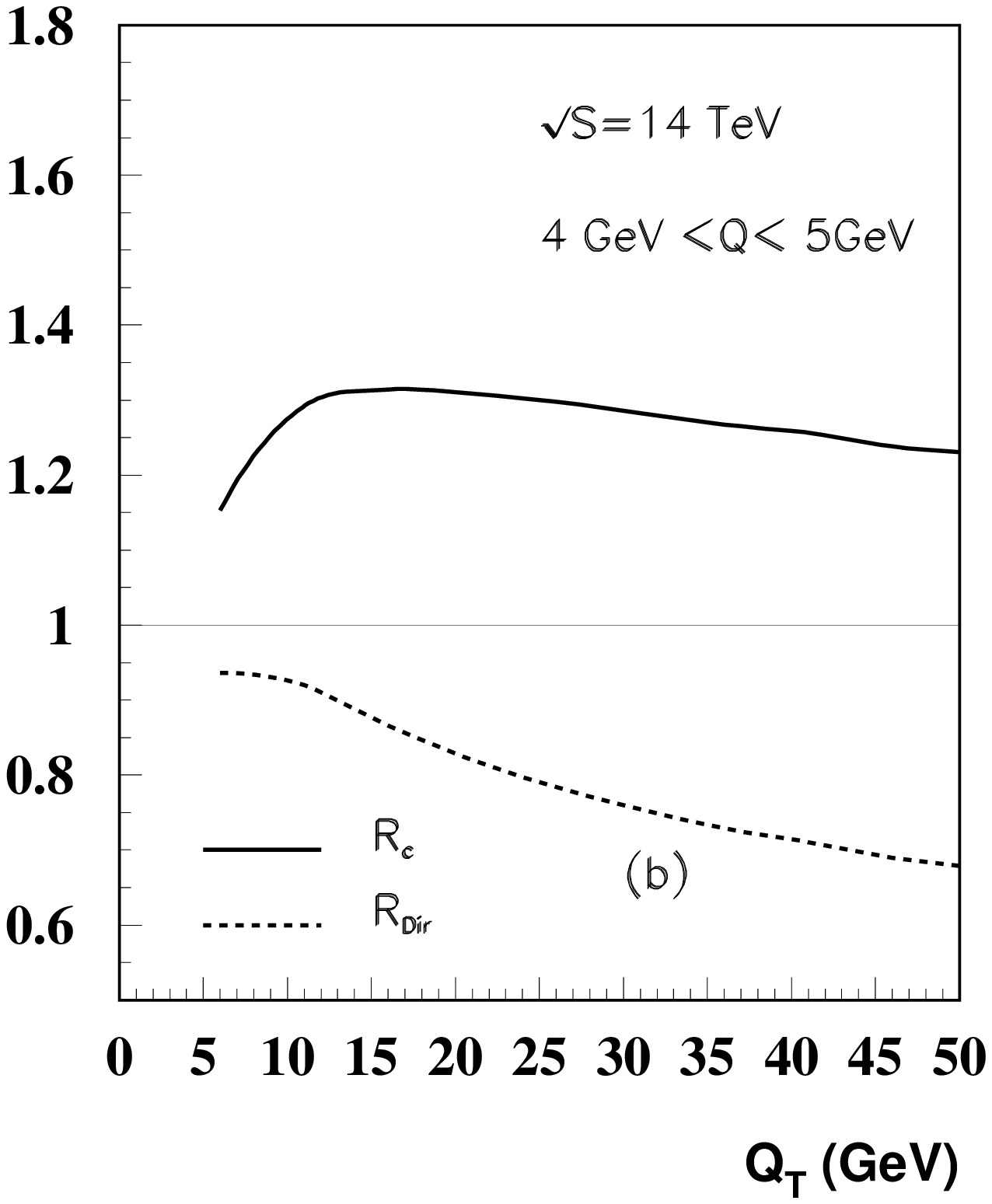,width=3.0in}
\end{minipage}
\end{center}
\caption{The ratios in Eqs.~(\protect\ref{R-con}) and
(\protect\ref{R-dir}) as a function of $Q_T$ at $y=0$ for the LHC
energy $\sqrt{S}=14$~TeV and mass intervals (a) $2\le Q\le 3$~GeV and 
(b) $4\le Q\le 5$~GeV.  Solid and dashed lines are for $R_c$ and $R_{Dir}$, 
respectively.} 
\label{fig15}
\end{figure}

We compare $R_{Dir}$ and $R_c$ in Figs.~\ref{fig14} and \ref{fig15}
at the Tevatron and LHC energies.  As expected, $R_{Dir}$ is 
smaller than $R_c$.  We remind the reader that in our notation, 
NLO corresponds to the pure $O(\alpha_s^2)$ contribution and 
LO corresponds to pure $O(\alpha_s)$.  Thus, the commonly defined 
$K$-factor is obtained by adding $1$ to the ratios 
Figs.~\ref{fig14} and \ref{fig15}.  The factor $K$ would be near 
$2.3$ for $Q_T \sim 10$ GeV 
(near $1.9$ for $Q_T \sim 50$ GeV) at $\sqrt S = 2$ TeV and 
$2\le Q\le 3$~GeV in the conventional approach, but somewhat 
smaller, $\sim 2$ and $\sim 1.6$, for our direct contribution.  The 
relatively smaller role of the NLO term lends greater confidence in the 
reliability of the calculated Drell-Yan cross sections in our modified 
formalism.

The denominators are identical in the definitions of the ratios 
$R_c$ and $R_{Dir}$ in Eqs.~(\ref{R-con}) and~(\ref{R-dir}).  The 
difference between the solid and dashed curves in Figs.~\ref{fig14} and 
\ref{fig15} is therefore the ratio of the logarithmic piece in NLO 
to the leading-order direct contribution.  The figures show that a 
considerable fraction of the NLO correction in conventional perturbative 
calculations comes from logarithmic contributions associated with 
bremsstrahlung radiation of the virtual photon.   As noted above, the NLO 
fragmentation contributions are not available yet, and, therefore, we do 
not show results here for $R_F$. 

Taken at face value, the ratio $R$ in Fig.~\ref{fig13} suggests that 
the cross section predicted in our modified formalism is modestly less 
than that in the conventional approach for $Q_T > Q$ and that the shape 
of the predicted $Q_T$ distribution is steeper.  However, we remind the 
reader that we have not included the NLO fragmentation contributions.  
When these are added in subsequent work, the predicted Drell-Yan cross 
sections from our modified factorization formula should be larger than those 
from the conventional NLO perturbative approach. Estimating the ratio of the
fragmentation contributions by the ratio of the direct contributions, 
$R_F \sim R_{Dir}$, we expect the Drell-Yan cross sections, with
the NLO fragmentation contributions included, will be greater by 10 to 30 \%
than the predicted totals in Figs.~\ref{fig7} -- \ref{fig9}.   For 
example, at $\sqrt S = 2$ TeV and $2\le Q\le 3$~GeV, we estimate 
increases of 15\% and 10\% at $Q_T = 9$ and $39$ GeV.  

\begin{figure}
\begin{center}
\begin{minipage}[t]{3in}
\epsfig{figure=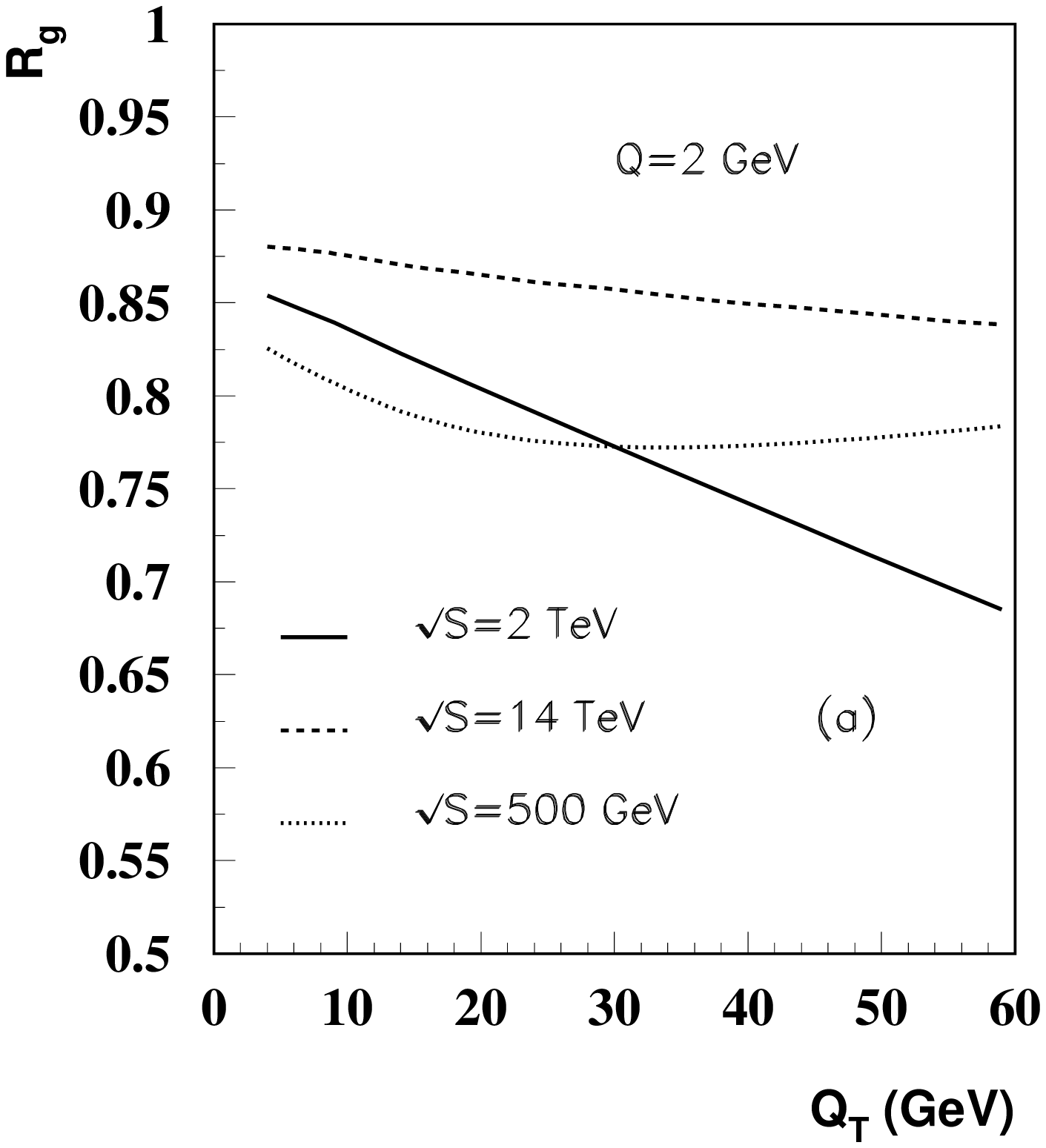,width=3.0in}
\end{minipage}
\hfill
\begin{minipage}[t]{3in}
\epsfig{figure=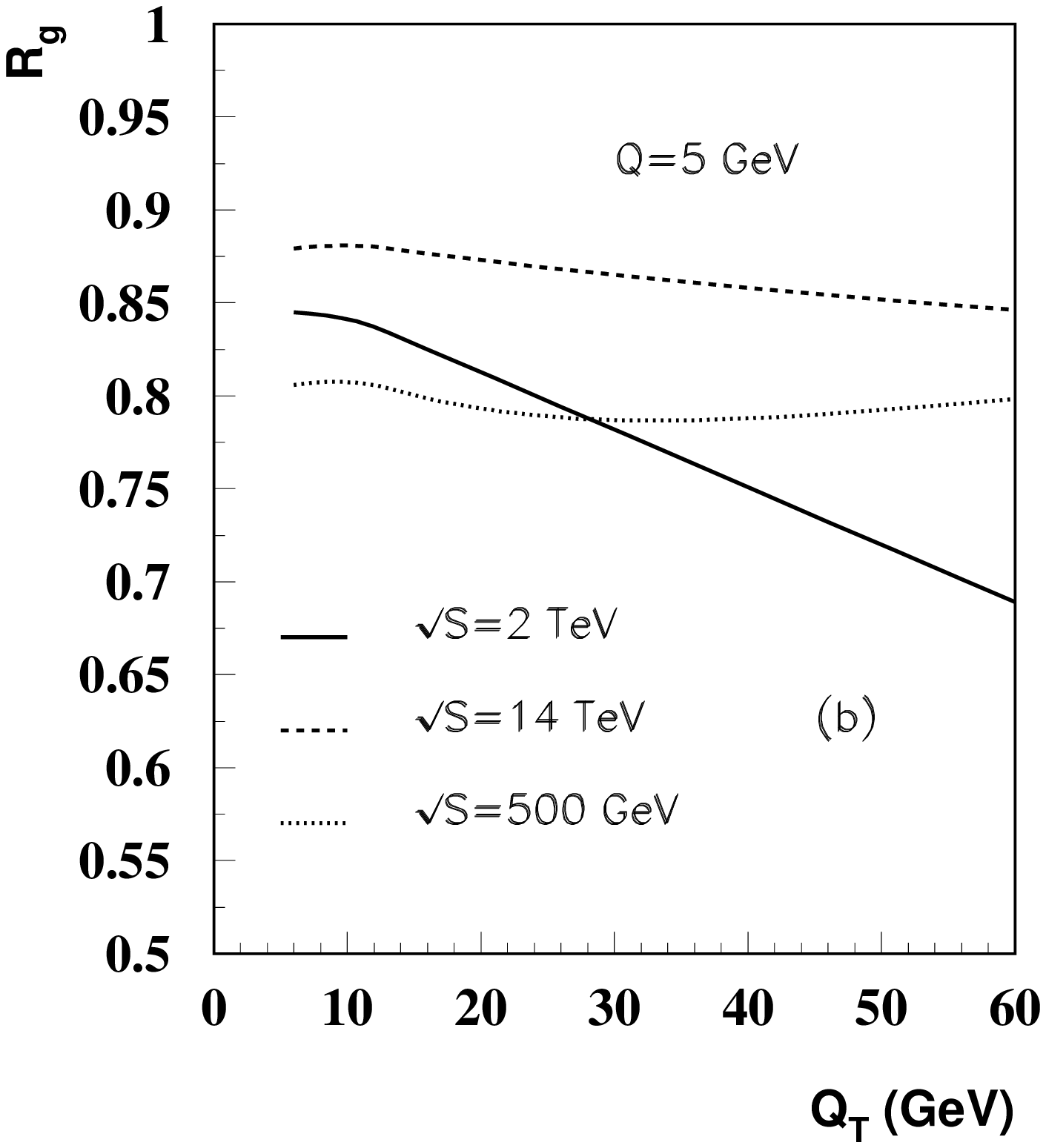,width=3.0in}
\end{minipage}
\end{center}
\caption{Ratio $R_g$ in Eq.~(\protect\ref{R-g}) as a function of $Q_T$
at $y=0$ at the Tevatron energy (solid), the LHC energy (dashed), and
the RHIC proton-proton energy (dotted) for virtual photon invariant 
masses (a) $Q=2$~GeV and (b) $Q=5$~GeV.}
\label{fig16}
\end{figure}

To demonstrate the relative size of gluon initiated contributions, we 
define the ratio
\begin{equation}
R_g = \left.
\frac{d{\sigma}_{AB\rightarrow \gamma^*(Q) X}(\mbox{gluon-initiated})}
     {dQ_T^2\,dy} \right/
\frac{d{\sigma}_{AB\rightarrow \gamma^*(Q) X}}{dQ_T^2\,dy}\, .
\label{R-g}
\end{equation}
The numerator includes the contributions from all partonic subprocesses with 
at least one initial-state gluon, and the denominator includes all subprocesses. 
In Fig.~\ref{fig16}, we show $R_g$ as a function of $Q_T$ at $y=0$ for the 
Tevatron energy (solid), the LHC energy (dashed), and the RHIC proton-proton 
energy (dotted) at $Q=2$~GeV and $Q=5$~GeV.  We confirm that gluon 
initiated subprocesses dominate the Drell-Yan cross section and that 
low mass Drell-Yan lepton-pair production at large transverse momentum 
is an excellent source of information on the gluon distribution~\cite{BGK-DY}.
The falloff of $R_g$ at very large $Q_T$ is related to the reduction of phase 
space and the fact that cross sections are evaluated at larger 
values of the partons' momentum fractions. In the Tevatron case,  
quark-antiquark subprocesses are expected to (re)assert their dominance 
at very large $Q_T$ because of the valence nature of the antiquark density 
in the incident antiproton.

%%%%%%%%%%%%%%%%%%%%%%%%%%%%%%%%%%%%%%%%%%%%%%%%%%%%%%%%%%%%%%%%%%%%%%%%%
\section{Conclusions}
\label{sec6}

In this paper we introduce a new factorization formalism 
for the Drell-Yan cross section at large $Q_T$.  It incorporates all-orders 
resummation of large logarithmic contributions into parton-to-virtual photon 
fragmentation functions. In addition to the logarithmic contributions, our 
expression for the cross section includes the usual 
{\em non}-logarithmic contributions, referred to often as {\em direct} 
contributions, in both leading $O(\alpha_s)$ and higher orders.  
This modified factorization formula agrees with the conventional fixed-order 
QCD expression when $Q_T^2\sim Q^2$.  The difference between the modified
factorization formula and the conventional formula is determined by QCD 
evolution of the virtual photon fragmentation functions.  Our modification 
reorganizes the {\it single} perturbative expansion of conventional QCD 
factorization into {\it two} expansions plus the perturbatively calculated
parton-to-virtual photon fragmentation functions.  The new expansions are 
evaluated at a single hard scale and are free of large $\ln^m(Q_T^2/Q^2)$ 
terms when $Q_T^2\gg Q^2$.  The ratio of the next-to-leading $O(\alpha_s^2)$ 
to leading order $O(\alpha_s)$ contributions is smaller in the new expansion 
than in the conventional case.  The reliability of the predictions is 
enhanced.   

As shown in Fig.~\ref{fig11}, the contributions from the logarithmic terms 
are important.  They can be as large as 70\% of the non-logarithmic contributions 
at the LHC energies.  At RHIC energies, the logarithmic contributions are 
smaller because phase space at large transverse momentum is limited and the 
penalty associated with large invariant mass $Q$ is felt more strongly.  
In Fig.~\ref{fig7}, \ref{fig8}, and \ref{fig9}, we present predictions 
through next-to-leading order at energies of interest for experiments at the 
Fermilab Tevatron, Brookhaven's RHIC, and the CERN Large Hadron Collider. 
Resummation somewhat steepens the shape of the predicted $Q_T$-spectrum relative 
to the conventional fixed-order approach.  However, at the order in perturbation 
theory at which we work, it has only a modest effect on the normalization.  We 
confirm that the Drell-Yan cross section at large $Q_T$ remains an excellent 
source of constraints on the gluon parton density.

%%%%%%%%%%%%%% Begin Acknowledgments %%%%%%%%%%%%%%%%%%%%%%%%%%%%%%%%

\section*{Acknowledgments}
Work in the High Energy Physics Division at Argonne National 
Laboratory is supported by the U.S. Department of Energy, 
Division of High Energy Physics, Contract W-31-109-ENG-38.
The research of J-W Q and X-F Z at Iowa State University is 
supported in part by the U.S. Department of Energy under Grant 
No. DE-FG02-87ER40731.

%%%%%%%%%%%%%% End Acknowledgments %%%%%%%%%%%%%%%%%%%%%%%%%%%%%

%%%%%%%%%%%%%%%%%%%%%%%%%%%%%%%%%%%%%%%%%%%%%%%%%%%%%%%%%%%%%%%%%%%%%%%%%

%%%%%%%%%%%%%% Begin References %%%%%%%%%%%%%%%%%%%%%%%%%%%%%%%%

%%%%%%%%%%%%%% Begin Figure Captions %%%%%%%%%%%%%%%%%%%%%%%%%%%%%%%%%%%

%%%%%%%%%%%%%%%%%%%%%%%%%%%%%%%%%%%%%%%%%%%%%%%%%%%%%%%%%%%%%%%%%%%%%%%%%
\end{document}